# Encoding Sensory and Motor Patterns as Time-Invariant Trajectories in Recurrent Neural Networks

**Authors:** Vishwa Goudar[2], Dean V. Buonomano[1,2,3,*]

**Affiliations:**
[1]Integrative Center for Learning & Memory, Departments of [2]Neurobiology and [3]Psychology, UCLA, Los Angeles, CA, 90095, USA.

**Correspondence:**
* Correspondence and requests for materials should be addressed to Dean V. Buonomano (dbuono@ucla.edu).

**Abstract:**

Much of the information the brain processes and stores is temporal in nature—a spoken word or a handwritten signature, for example, is defined by how it unfolds in time. However, it remains unclear how neural circuits encode complex time-varying patterns. We show that by tuning the weights of a recurrent neural network (RNN), it can recognize and then transcribe spoken digits. The model elucidates how neural dynamics in cortical networks may resolve three fundamental challenges: first, encode multiple time-varying sensory *and* motor patterns as stable neural trajectories; second, generalize across relevant spatial features; third, identify the same stimuli played at different speeds—we show that this temporal invariance emerges because the recurrent dynamics generate neural trajectories with appropriately modulated angular velocities. Together our results generate testable predictions as to how recurrent networks may use different mechanisms to generalize across the relevant spatial and temporal features of complex time-varying stimuli.

## Introduction:

Many, if not most, of the tasks the brain performs are inherently temporal in nature: from recognizing and generating complex spatiotemporal patterns—such as a phoneme sequence that composes a word—to creating temporal expectations of when an event will occur (Mauk and Buonomano, 2004; Nobre et al., 2007; Ivry and Schlerf, 2008; Merchant et al., 2013; Hopfield, 2015). In contrast to the representation of an object in a static image, such as a picture of a face, robust decoding and encoding of a time-varying patterns must rely on the spatiotemporal dependencies inherent to the pattern. How does the brain accomplish this? One highly influential theory in neuroscience holds that information is stored as fixed-point attractors that emerge in the dynamic activity of the brain's recurrently connected circuits (Hopfield, 1982; Hopfield and Tank, 1986; Amit and Brunel, 1997; Wang, 2001). Two limitations of this framework are: 1) It fails to capture the temporal aspect of stored information, thus forcing many computational models to "spatialize" time—that is, they treat the temporal component of time-varying patterns as additional spatial dimensions (Rabiner, 1989; Waibel et al., 1989; Elman, 1990; Hinton et al., 2012; Mnih et al., 2015); 2) it does not capture a fundamental feature of how the brain processes temporal information: temporal invariance. For example, humans readily recognize temporally warped—compressed or dilated—speech or music.

While the mechanisms that underlie the brain's ability to perform a broad range of spatiotemporal tasks in the sensory and motor domains are not known, there is mounting theoretical and experimental evidence that our ability to tell time on the sub-second scale, and represent time-varying patterns, relies on the inherent *continuous* dynamics, and computational potential, of recurrent neural networks (Rabinovich et al., 2008; Buonomano and Maass, 2009; Mante et al., 2013; Crowe et al., 2014; Carnevale et al., 2015; Li et al., 2016). This framework alleviates the restrictions imposed by traditional discrete-time or fixed-point attractor models, by relying on the recurrent connections to implicitly maintain an ongoing memory of the pattern. Indeed, recent computational studies have established that by tuning the weights within recurrent neural network (RNN) models, it is possible to *robustly* store and generate complex time-varying motor patterns (Laje and Buonomano, 2013; Abbott et al., 2016; Rajan et al., 2016). What remains unknown is whether the same approach can be used to robustly discriminate time-varying sensory patterns. Indeed, this poses a challenging problem because in order to effectively process spatiotemporal stimuli the dynamics of an RNN must be sensitive to the relevant spatial and temporal features of the sensory stimuli, while being able to generalize across their natural spatial and temporal variations.

Neuroscientists have typically distinguished between sensory and motor areas; but it is well established that activity in sensory areas is strongly influenced by motor behavior, and that sensory stimuli can modulate activity in motor areas—furthermore, some brain areas are characterized as being sensorimotor (Doupe and Kuhl, 1999; Ayaz et al., 2013; Chang et al., 2013; Schneider and Mooney, 2015; Cheung et al., 2016). Computationally, sensory and motor processing are understood to have disparate requirements—during sensory processing, network dynamics should primarily be driven by the sensory inputs; in contrast, during motor processing neural dynamics should be autonomous and driven primarily by recurrent interactions. It remains unclear how a single network could accomplish both tasks, and couple them when necessary. Indeed, to date, no previous models have shown that the same network can satisfy the requirements for both sensory and motor processing. Here we show that the same RNN can reliably function in both a sensory and motor regime. Specifically the same RNN can convert complex time-varying sensory patterns into motor patterns—thus performing a transcription task in which spoken digits are identified and read out as "handwritten" digits.

In the temporal domain, understanding how the brain recognizes temporally compressed or dilated patterns represents a long-standing challenge. This temporal invariance is particularly evident in our ability to recognize temporally warped speech or music (Miller et al., 1984; Sebastián-Gallés et al., 2000). Our results suggest that one advantage of storing time-varying patterns as neural trajectories within RNNs is that, under the appropriate conditions, this strategy naturally accounts for temporal invariance.

**Results:**

We first asked if a single RNN could perform a complex sensory-motor task: transcribing spoken digits into handwritten digits. A *continuous-time* firing-rate RNN with randomly assigned sparse recurrent connections was used (Sompolinsky et al., 1988a). The strengths of these recurrent connections were initialized to be relatively strong; thus, before training the network was in a so-called high-gain regime ($g$ = 1.6) that is characterized by self-perpetuating and chaotic activity (Materials and Methods). The transcription task is divided into a sensory and a motor epoch. During the sensory epoch, the RNN is presented with a spoken digit, and over the ensuing motor epoch the RNN drives an output pattern transcribing the presented digit via three motor outputs $x$, $y$, and $z$—activity in the $x$ and $y$ units determines the 2D coordinates of a “pen on paper”, while $z$ determines if the “pen” is in contact with the “paper” or not. **Figure 1A** illustrates the network architecture and transcription task for the digit "2". Successful performance of this transcription task requires that the RNN: 1) encode spoken digits with a set of neural trajectories in high-dimensional phase space; 2) autonomously generate digit-specific high-dimensional trajectories during the motor epoch, in order to drive the digit-specific output patterns; and most importantly 3) generate trajectories that are stable so they may encode each digit’s sensory pattern, sensory-motor transition, and motor pattern in a manner that is invariant not just to background noise but also to spatiotemporal variations of the spoken digits, including temporal warping.

We attempted to satisfy these three conditions by training the RNN using a supervised learning rule, in which the recurrent units were trained to robustly reproduce their "innate" patterns of activity (Laje and Buonomano, 2013)—that is, those generated in the untrained network—using the recursive-least-squares learning rule (Haykin, 2002; Sussillo and Abbott, 2009). In essence, the RNN is trained to reproduce one digit-specific “innate” trajectory in response to all training utterances of a given digit. For example, the pattern of activity produced in response to a "template" utterance of a given digit from the beginning of the sensory epoch to the end of the motor epoch is taken as the innate trajectory; the network is then trained to reproduce this trajectory in response to other utterances of the same digit. Furthermore, it is trained to do so regardless of the initial state of the network’s units, and in the presence of continuous background noise (Materials and Methods). Only after training the RNN (*recurrent training*), are the output units trained to generate the handwriting patterns representing each of the digits during the motor epoch (*output training*)—this separation of the recurrent and output training phases allows for rapid training and retraining of the outputs to produce arbitrary motor patterns, without retraining the recurrent weights. The output training phase is performed using standard supervised methods (Materials and Methods). **Figure 1B** shows the outputs of a trained network cross-tested on ten digits (0-9) across five speakers. Following training, the network successfully transcribes the utterances used during training (marked by asterisks). More importantly, performance generalizes to novel utterances and speakers in the dataset despite significant variations in the duration and spatiotemporal structure across utterances (**Supplementary Movie 1**).

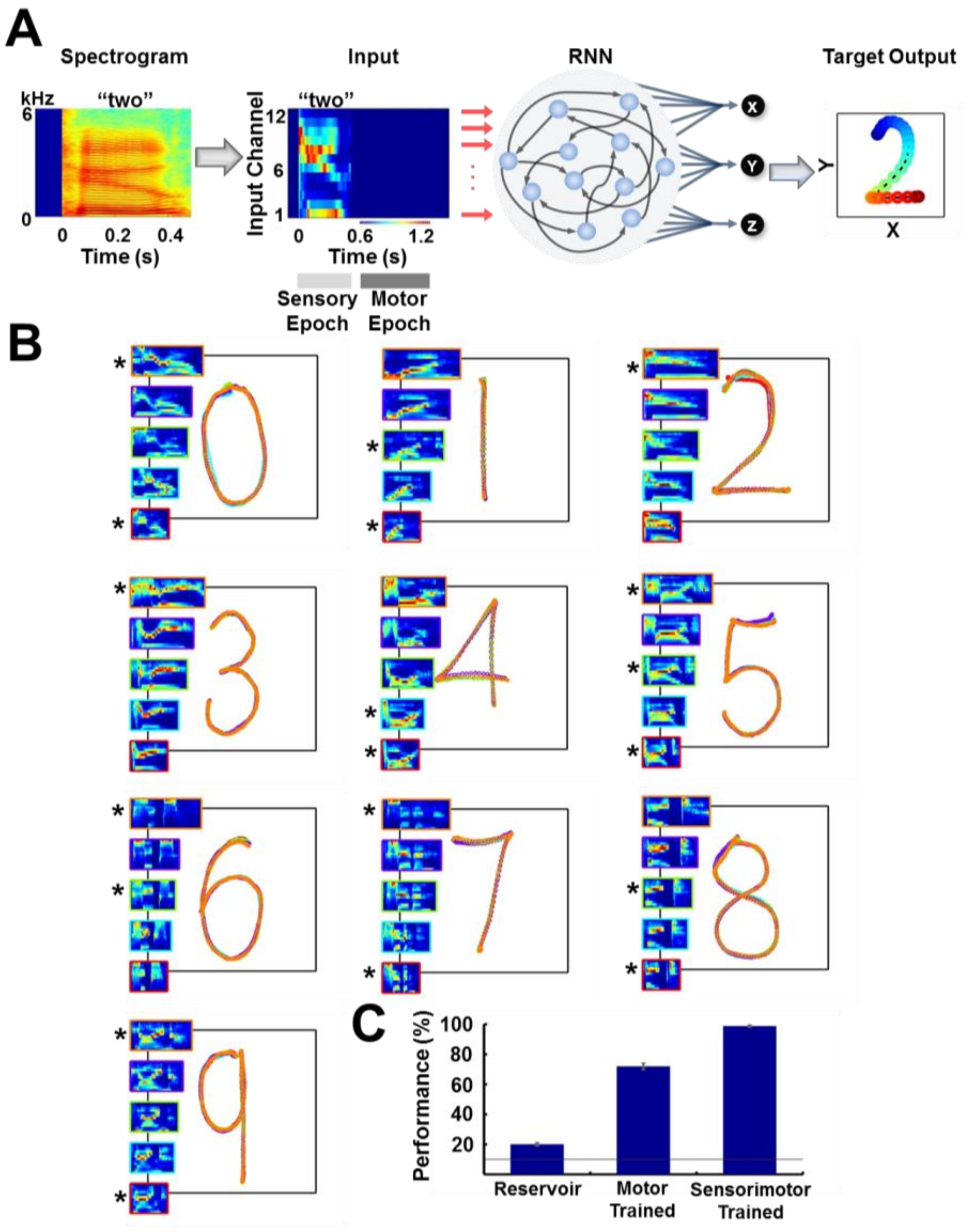

**Figure 1: Trained RNNs perform a sensorimotor spoken-to-handwritten digit transcription task.**
**(A)** Transcription task. The spectrogram of a spoken digit, e.g. "two", is transformed to a 12-channel cochleogram that serves as the continuous-time input to a RNN during the sensory epoch of each trial. During the motor epoch, the output units must transform the high-dimensional RNN activity into a low-dimensional "handwritten" motor pattern corresponding to the spoken digit (the z output unit indicates whether the pen is in contact with the "paper"). The colors of the output pattern (right panel) represent time (as defined in the "Input" panel). **(B)** Overlaid outputs of a trained RNN (*N*=4000) for 5 sample utterances of each of 10 digits. For all digits, each output pattern is color-coded to the bounding box of the corresponding cochleogram (inset). Sample utterances shown are a mix of trained (*) and novel utterances, and span the range of utterance durations in the dataset. **(C)** Transcription performance of three different types of RNNs on novel utterances. Performance was based on images of the output as classified by a deep CNN handwritten digit classifier. The control groups include untrained RNNs ("reservoir") and RNNs trained only during the motor epoch (i.e., just to reproduce the handwritten patterns; see Supplementary Fig. 1). Output unit training was performed identically for all networks. Bars represent mean values over 3 replications, and error bars indicate standard errors of the mean. Line indicates chance performance (10%). The RNNs were trained on 90 utterances (10 digits, 3 subjects, 3 utterances per subject per digit). They were then tested on 410 novel utterances (across five speakers, including two novel speakers), with 10 test trials per utterance. $I_0$ was set to 0.5 during network training (if applicable), and to 0.05 during output training and testing.

To quantify performance, we need an objective measure of the quality of the motor output for each test utterance. This itself represents a classification task, wherein images of RNN-generated transcriptions must be assigned to one of the ten possible digits. Rather than use a human-based performance measure, we used a standard deep convolutional neural network (CNN) (LeCun et al., 2015) to rate the performance of trained RNNs (Materials and Methods)—i.e., the CNN was used to determine if each "handwritten" digit was correct or not. Performance of trained RNNs was 98.7% on a test set of 410 novel stimuli (**Fig. 1C**). Since untrained RNNs ("reservoir networks") are in and of themselves capable of performing many interesting computations (Maass et al., 2002; Jaeger and Haas, 2004; Lukoševičius and Jaeger, 2009), it is important to determine how much of this performance is dependent on the training of the RNNs. We thus examined a control group wherein the three output units are trained as they were for the trained RNNs, but the RNN itself was not trained (the "reservoir"). The performance was very poor (20%), in large part because during the motor epoch a reservoir RNN is operating in an autonomous mode that is chaotic—making it difficult for the output units to learn to produce the target output patterns. We also examined the effects of training an RNN only during the motor epoch, resulting in a performance of 71%. This significant improvement over the reservoir networks is a consequence of the fact that a network driven by external inputs can encode sensory stimuli despite a lack of sensory epoch training, due to the stabilizing influence of the external inputs (see below). Yet, despite the improvement in performance, some digits were almost always misclassified (**Supplementary Fig. 1**). The difference in performance between the trained RNNs (sensory and motor training) and the exclusively motor trained RNNs, confirms the importance of tuning the recurrent weights to the sensory discrimination component of the task (see below).

***Stability of the Neural Trajectories.***

RNNs operating in high-gain regimes are prone to chaotic behavior (Sompolinsky et al., 1988b; Rajan et al., 2010b), because the strong recurrent feedback characteristics of these networks rapidly amplify any noise or perturbations. It is thus critical to demonstrate that the above performance is robust to background noise and perturbations during the sensory and motor epochs. The distinction between epochs is critical since the RNN is operating in fundamentally different regimes during the sensory and motor epochs. During the sensory epoch the recurrently generated internal dynamics are partially suppressed (or "clamped") as a result of the external input, yet during the motor epoch all activity is internally generated. To examine the stability of these sensory and motor "object" representations, we briefly perturbed the internal dynamics of the RNN during either the sensory or motor epochs. Sensory epoch perturbations were introduced half way through the presentation of each utterance, while motor epoch perturbations were introduced at the 10% mark of the motor epoch. **Figure 2A** shows the activity of a hundred units from a trained RNN (and the resulting output), when it is presented with the digit “three” and strongly perturbed (amplitude = 2) in the motor epoch. The resulting transcription briefly deviates from an unperturbed one (grey backdrop), but recovers mid-voyage. The sensory epoch is less sensitive to perturbations—as would be expected because the presence of the external input serves as a stabilizing influence (Rajan et al., 2010b). During the motor epoch the RNN is operating autonomously as a "dynamic attractor", that is, once bumped off its trajectory it maintains a memory of its current voyage and is able to return to the original trajectory (Laje and Buonomano, 2013) (**Fig. 2B**). Performance measurements using the CNN classifier reveal a graceful degradation across a wide range of perturbation magnitudes, and confirm the superior robustness of the sensory epoch (**Fig. 2C**).

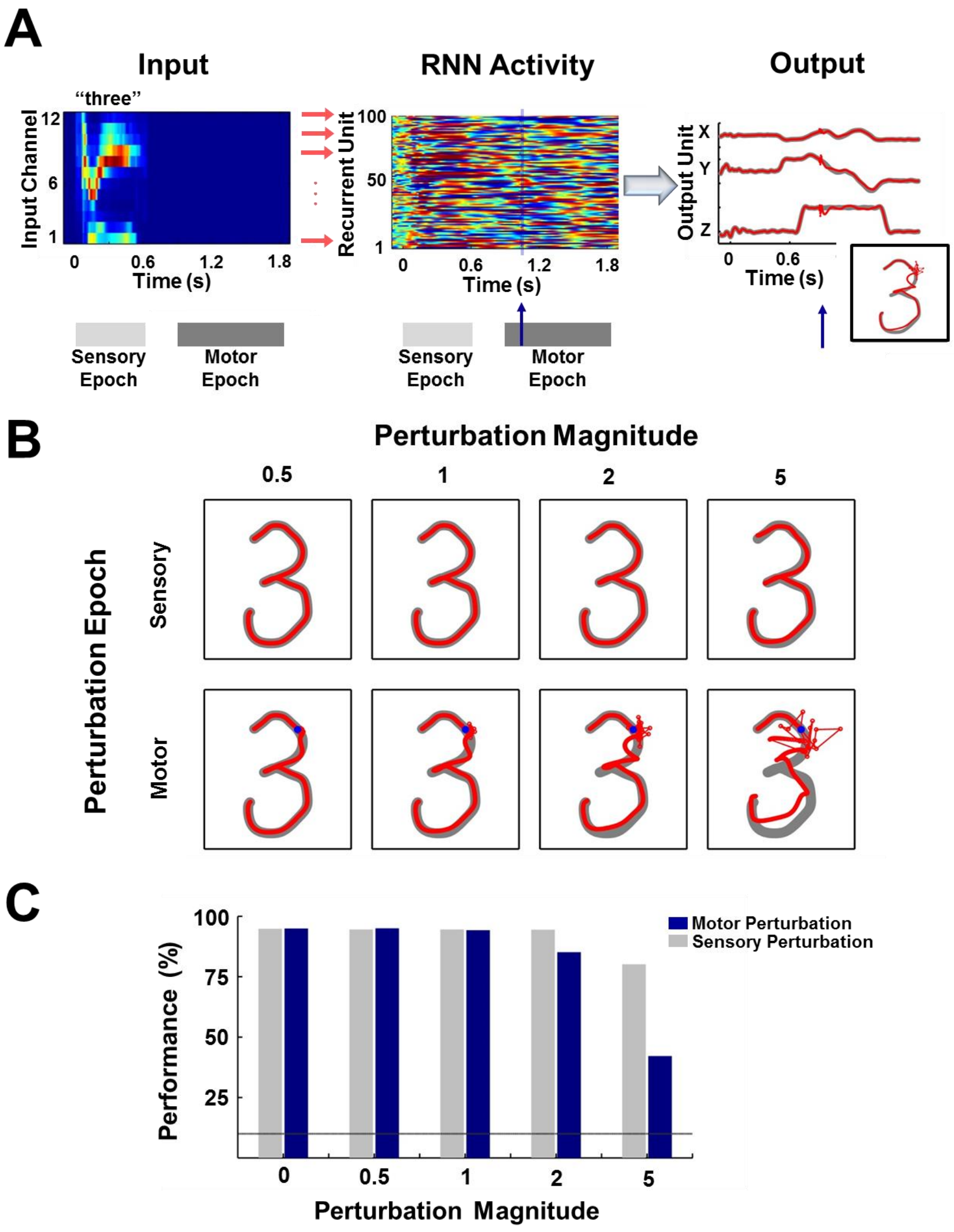

**Figure 2: Digit transcription is robust to perturbations during the sensory and motor epochs.**
**(A)** Schematic of a perturbation experiment. The motor trajectory of a trained RNN (*N*=2100; 100 sample units shown) for the spoken digit "three", is perturbed with a 25 ms pulse (amplitude = 2). The pulse (blue arrow) causes a disruption of the network trajectory, but the trajectory quickly recovers and returns to the dynamic attractor, as is evident from the plots on the right comparing the output unit values and the transcribed pattern for a trial with (red forefront) and without (gray backdrop) the perturbation. **(B)** Sample motor patterns generated by the network in response to perturbations of increasing magnitude, applied either during the sensory or motor epochs. Sensory epoch perturbations were applied halfway into the epoch, while motor epoch perturbations were applied at the 10% mark of the epoch (indicated by a blue dot). **(C)** Impact of perturbations on transcription performance (measured by the deep CNN classifier) for test utterances. Bars represent mean performance over ten trials, with a different (randomly selected) perturbation pulse applied at each trial. Line indicates chance performance. The performance measures establish that the encoding trajectories are stable to background noise perturbations, with transcription performance degrading gracefully as the perturbation magnitude increases. Furthermore, at all perturbation magnitudes, the sensory encodings are more robust than their motor counterparts due to the suppressive effects of the sensory input. The network was trained on 30 utterances (3 utterances of each digit by 1 subject) and tested on 70 (7 utterances of each digit by 1 subject). $I_0$ was set to 0.25 during network training, to 0.01 during output training, and to 0 during testing except for the duration of the perturbation pulse.

***RNN Training Sculpts Network Dynamics to Enhance Discrimination.***

How does tuning the recurrent weights allow the RNN to stably encode both sensory and motor information in neural trajectories, despite considerable spatiotemporal differences between digit utterances (**Fig. 1B**, insets)? Firing rate traces of sample units in a trained RNN show more similar patterns of activity in response to different utterances of the same digit, when compared to a reservoir RNN (**Fig. 3A-B**). To better examine the structure of the population representations, we can visualize and compare the neural trajectories in response to multiple utterances (learned and novel) of the same digit, during the sensory and motor epochs, in principal component analysis (PCA) subspace. The trajectories produced in the untrained network by utterances of the digits "six" and "eight", during the sensory epoch, occupied a fairly large abutting volume of PCA subspace (**Fig. 3C**); and during the motor epoch the trajectories were highly variable—the network strongly and perpetually amplifies the differences between utterances, because it is in a chaotic regime. In contrast, in the trained RNN, sensory epoch trajectories of each digit were restricted to a narrower volume, and better separated from the volume representing the other digit (**Fig. 3D**). During the motor epoch, the trajectories for different utterances of the same digit were constrained to a much narrower tube—reflecting the dynamic attractor—and better separated from the motor trajectories of the other digit. The set of all within-digit trajectories can be thought of as populating a hypertube that delimits the volume of phase space traversed by the digit. During the sensory epoch, this hypertube of trajectories represents a memory of a family of related spatiotemporal objects: the spoken digit. During the motor epoch the network is autonomous, and the hypertube of all within-digit trajectories narrows to a dynamic attractor that represents the “motor memory” of the corresponding handwritten digit.

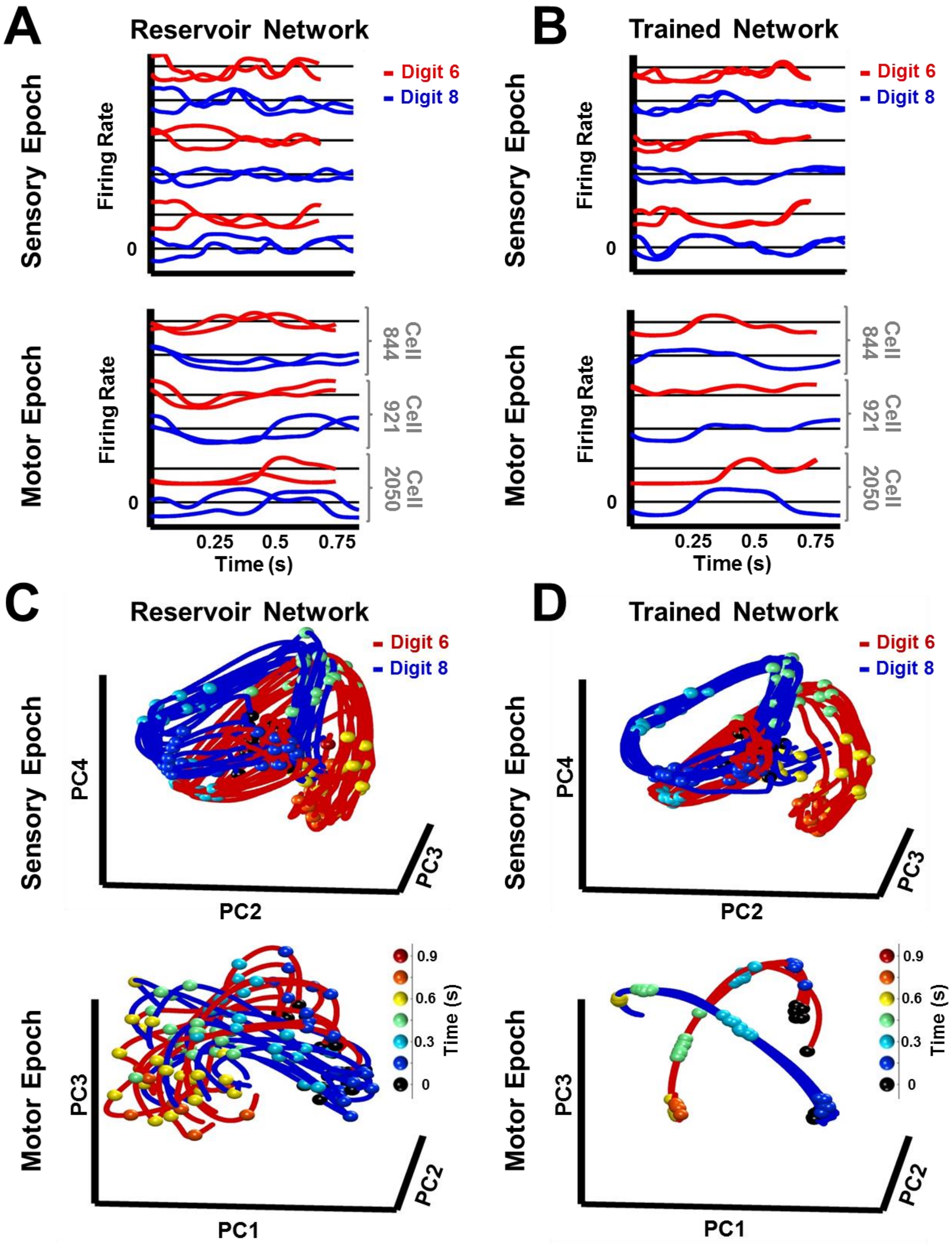

**Figure 3: Trained RNNs generate convergent continuous neural trajectories in response to different instances of the same spatiotemporal object.**
**(A-B)** Neural activity patterns of three sample units of a reservoir **(A)** and trained **(B)** network, in response to a trained and a novel utterance each of the digits “six” (red traces) and “eight” (blue traces) during the sensory (top) and motor (bottom) epochs. Utterances of similar duration were chosen for each digit, to allow for a direct comparison, without temporal warping, of the corresponding pair of sensory epoch traces. **(C-D)** Projections in PCA space of the sensory and motor trajectories for 10 utterances each of the digits “six” and “eight” generated by the reservoir **(C)** and trained **(D)** networks. Colored spheres represent time intervals of 150 ms. Compared to the reservoir network, in the trained RNN the activity patterns in response to different utterances of the same digit are closer, and the trajectories in response to different digits are better separated. Both networks were composed of 2100 units. Network training was performed with 30 utterances (1 subject, 10 digits, 3 utterances per digit). $I_0$ was set to 0.5 during network training, and to 0 while recording trajectories for the analysis.

It is well established that cortical circuits undergo experience-dependent plasticity—a process that seems to result in the optimization or specialization of those circuits to the tasks the animal is exposed to (Buonomano and Merzenich, 1998; Crist et al., 2001; Feldman and Brecht, 2005; Karmarkar and Dan, 2006). The above results establish that training the RNN does improve discrimination performance—the recurrent weights are tuned to the task at hand—but leaves open the question of how exactly this is accomplished. To answer this question we asked if the Euclidean distances between trajectories in response to different utterances of the same digit (within-digit distance), and utterances of different digits (between-digit distance), were significantly altered in comparison to the reservoir (untrained) RNN. Training significantly decreases the mean within-digit distances during the sensory epoch (**Fig. 4A**). Importantly, in doing so, it does not diminish the large separations between trajectories of different digits. The same effect, much enhanced, is observed during the motor epoch, a consequence of the formation of dynamic attractors. Successful formation of these attractors is strongly influenced by an unambiguous and stable sensory epoch-motor epoch transition of the network dynamics. This transition relies on the high between-digit separation and low within-digit separation of sensory epoch trajectories, particularly at the end of the sensory epoch—more convergent within-digit sensory encodings result in more stereotyped initial conditions for the motor trajectories, in turn, resulting in the reliable formation of and transition to motor pattern-encoding dynamic attractors.

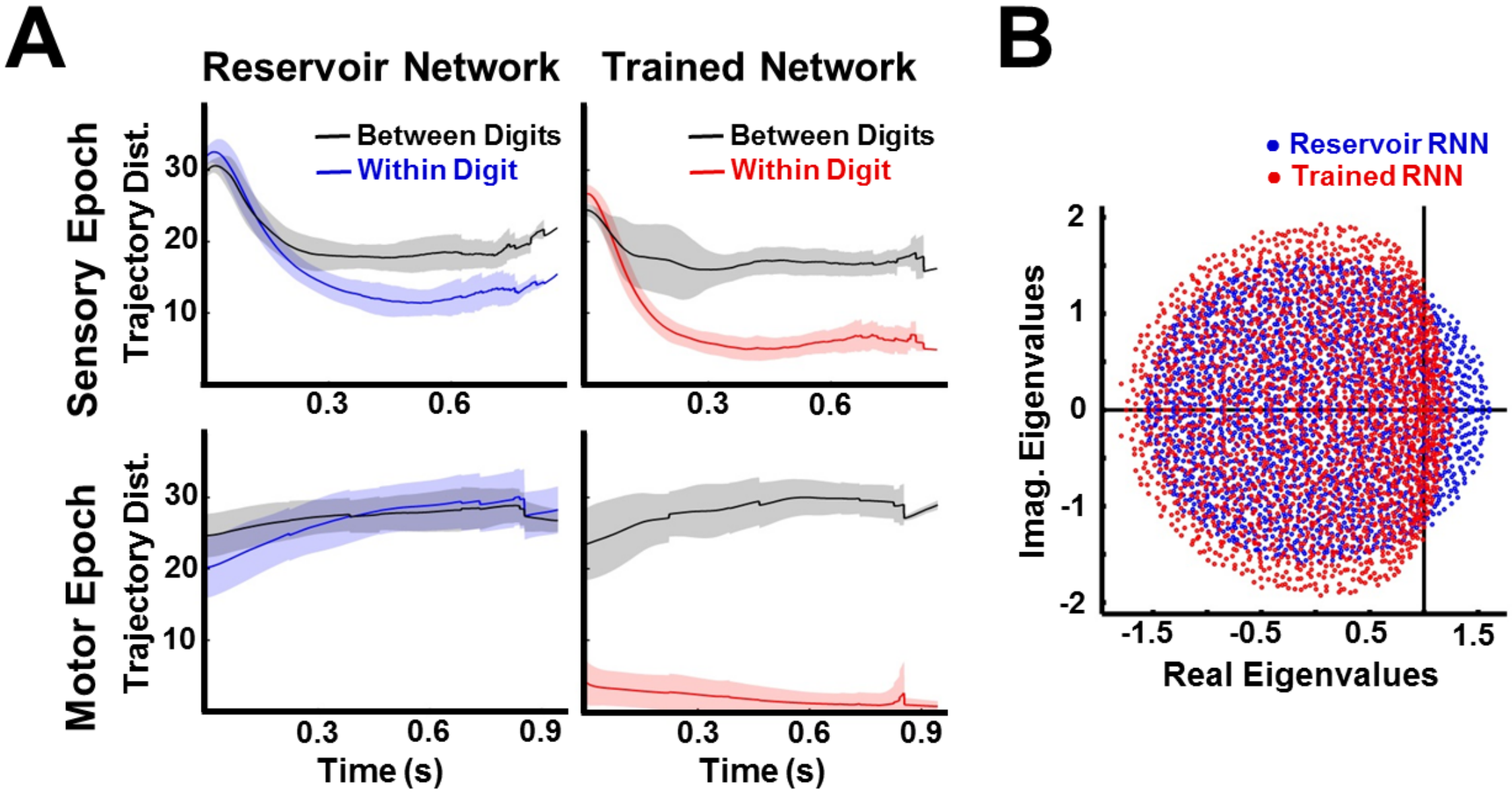

**Figure 4: Trained RNNs encode both sensory and motor objects as well separated neural trajectories.** **(A)** Euclidean distance between trajectories of the same digit (within-digit) versus those of different digits (between-digit). At each time step, the trajectory distances represent the mean and SD (shading) over pairs of one hundred utterances (1 subject, 10 digits, 10 utterances per digit). During the sensory epoch, training brings trajectories for the same digit closer together, while maintaining a large separation between trajectories of different digits, thereby improving discriminability. A similar, but stronger, effect is observed during the motor epoch. **(B)** Comparison of the eigenspectrum of the recurrent weight matrix ($\boldsymbol{W}^{\boldsymbol{R}}$) in a reservoir and trained network. Both networks were composed of 2100 units. Network training was performed with 30 utterances (1 subject, 10 digits, 3 utterances per digit). $I_0$ was set to 0.5 during network training, and to 0 while recording trajectories for the analysis.

The finding that training sculpts the dynamics of the RNN during the sensory epoch is also a critical one because it shows that even though RNN activity is governed in part by the external input, the internal connections are critical in that tuning them effectively improves the interaction between the sensory input and the internally generated dynamics. This improvement is expressed as a collapsing of the family of trajectories representing a digit into a narrower hypertube. These results follow as a direct consequence of the supervised recurrent training paradigm (Materials and Methods)—a single innate trajectory serves as a common target for different training utterances of a digit, thereby inducing changes in the network's recurrent dynamics that are necessary to encode the different utterances along similar neural trajectories; However, different digits are encoded in rapidly divergent trajectories because the target trajectories for each digit are generated by an untrained chaotic network. Further evidence that training dramatically influences the weight matrix of the RNN is demonstrated by the change in its eigenspectrum (**Fig. 4B**). Because the network is initialized with normally distributed weights with a gain of 1.6 (Materials and Methods), the eigenspectrum of the reservoir's weight matrix lies in a circle of radius approximately 1.6. Training results in a compression of those eigenvalues with real part larger than one, bringing the maximal real part of the eigenvalues closer to one. In a linear network, when all eigenvalues have a real part less than one, it implies that the network's activity will decay to zero when it operates autonomously; however our network is nonlinear and does not operate exclusively in the autonomous mode, so interpretations of the eigenspectrum of the weight matrix are not straightforward—nevertheless the compression of the eigenspectrum is consistent with the increased stability of the network during the sensory and motor epochs (Rajan and Abbott, 2006; Ostojic, 2014).

### ***Balance between Recurrent and Input Dynamics is Crucial for Discrimination.***

The above results provide insights as to how a single neural circuit can function in two seemingly distinct computational modes: sensory and motor. Sensory discrimination requires circuits to be highly responsive to external stimuli in order to categorize inputs into discrete classes. In contrast, motor tasks require autonomous generation of spatiotemporal patterns, and thus need the network dynamics to be somewhat resistant to external inputs. In the network described here, this balance depends on the recurrent weights and the magnitude of the external drive.

The recurrent weights must be strong—that is the network must be in a high-gain regime—for two reasons. First, network dynamics in high-gain regimes are high-dimensional (Rajan et al., 2010a), which naturally yields well-separated trajectories for different digits. A goal of training then, is to achieve recurrent suppression of within-digit separation, while retaining the large between-digit separation (**Fig. 4**). Second, as noted earlier, a network operating in a high-gain regime is inherently capable of generating the self-perpetuating activity required in the motor epoch, which is then stabilized by the training procedure. Thus during the sensory epoch, the RNN is operating in a regime that is both sensitive to external input and influenced by strong internal recurrent dynamics. In the motor mode, the RNN operates autonomously, generating locally stable trajectories that serve as a high-dimensional “engine” that drives arbitrary low dimensional output patterns.

The ability of the network to meaningfully process input patterns during the sensory epoch, so that they are easy to discriminate, also depends on the strength of the inputs. To better understand the impact of input amplitude on the encoding trajectories and their discriminability, before and after training, we parametrically varied the input amplitude and studied the resulting RNN dynamics during the sensory epoch. In the reservoir, both within and between-digit distances diminish as the input amplitude increases (**Fig. 5A, left panel**), consistent with a previous study showing that strong inputs can progressively dominate and override the internal dynamics of an RNN (Rajan et al., 2010b). In the extreme, input-dominated regimes effectively void the recurrent weights and render training useless (**Fig. 5A, right panel**). The poor discriminability at high input amplitudes is further confirmed by dimensionality measurements of sensory epoch trajectories (Materials and Methods), which mirror the low-dimensionality of the input cochleograms, regardless of training (**Fig. 5B**). In contrast, low-input amplitude regimes are dominated by the RNN’s internal dynamics. Reservoirs in this regime produce chaotic high-dimensional dynamics and are insufficiently sensitivity to external inputs; thus the within- and between-digit distances are similarly high (**Fig. 5A, left**). Training strongly alters these dynamics, lowering its dimensionality (**Fig. 5B**). However, as a consequence of poor input-sensitivity, these changes fail to improve discrimination—trained RNNs in this regime still sustain similar within and between-digit distances (**Fig. 5B, right**). RNNs trained at intermediate input amplitudes (0.5, 5) are both sensitive to the input and able to discriminate between digits, because their sensory epoch trajectories are shaped both by the input and internal dynamics. It is thus critical that the input drive be strong enough to influence ongoing activity but not strong enough to "erase" recent information encoded in the current trajectory.

***Mechanisms Underlying Spatial (Spectral) Generalization.***

The attracting dynamics of the network result in each digit being represented in a narrow hypertube. These representations must be robust across both spectral and temporal (changes in speed) differences in the utterances of each digit. Next, we separately examine the mechanisms underlying spatial and temporal generalization. Variation in the spectral structure

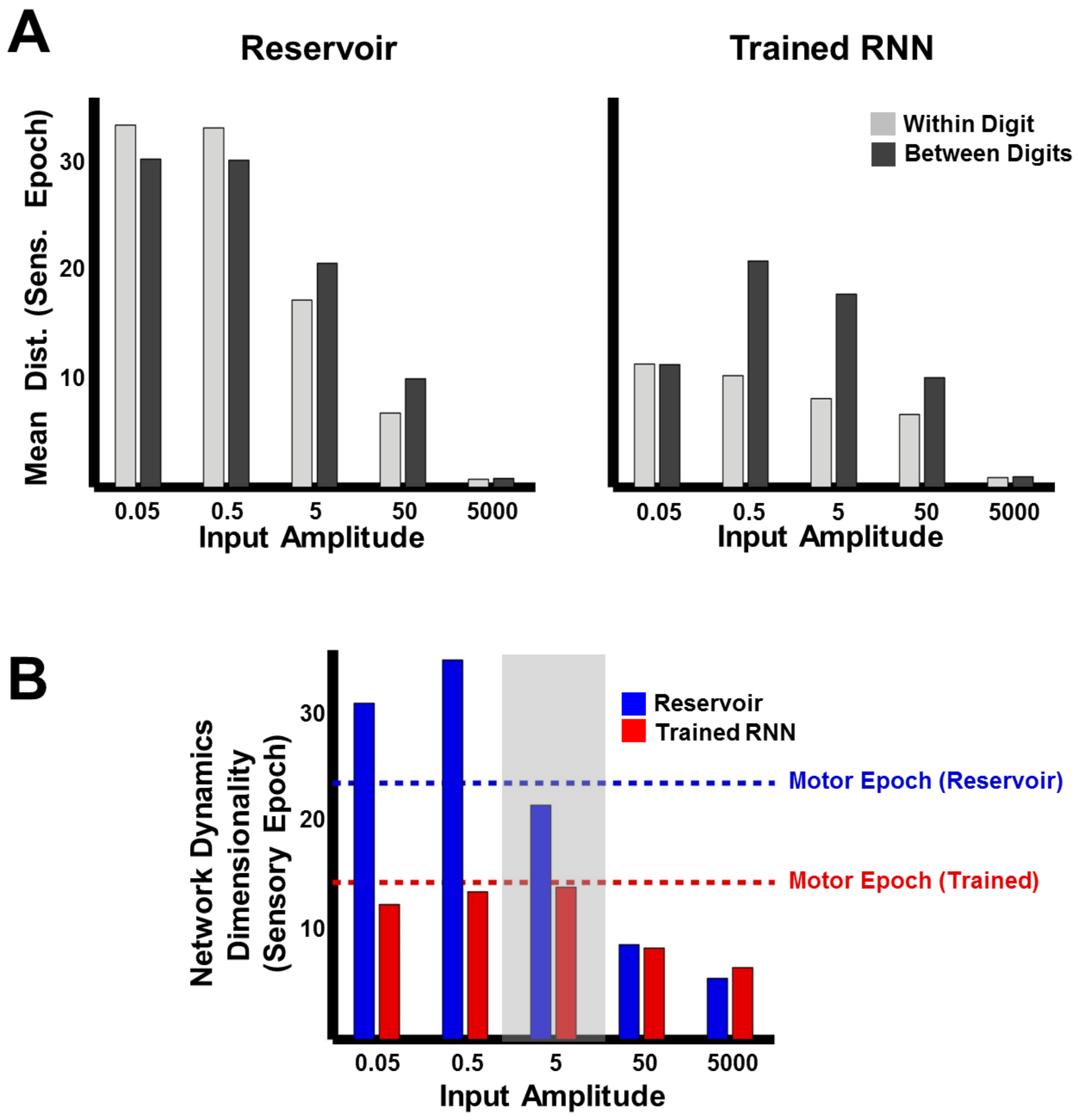

**Figure 5: Trajectory separation in reservoir and trained RNNs as a function of input amplitude.**
**(A)** Comparison of mean within- and between-digit distances of the sensory epoch trajectories in reservoir and trained networks (N=2100) at different input amplitudes. Bars represent mean of the time-averaged distances for all utterances of all digits (1 subject, 10 digits, 10 utterances per digit). **(B)** Dimensionality of sensory epoch trajectories in the reservoir and trained networks at different input amplitudes. Simulations, including training, in all other figures were performed at an input amplitude of 5 (gray highlight). Dashed lines indicate the dimensionality of the motor epoch trajectories in the reservoir (blue) and trained (red) networks, when the input amplitude was 5. Network training was performed with 30 utterances (1 subject, 10 digits, 3 utterances per digit). The networks were trained with $I_0$ set to 0.25, and trajectories were recorded with $I_0$ set to 0.

of digits—spectral noise" (**Fig. 6A**)—is qualitatively different from background noise. First, spectral noise exhibits time-dependent structured relationships with the input signal (they may be correlated, anti-correlated or tend to occur along specific directions during specific parts of the input). For example, the spectral differences between utterances of one phoneme within a digit will be different than those of another. Invariance then requires the network to isolate and integrate the signal while suppressing signal-dependent spectral noise. Second, spectral noise is a form of correlated noise. The noise in each input channel is simultaneously delivered

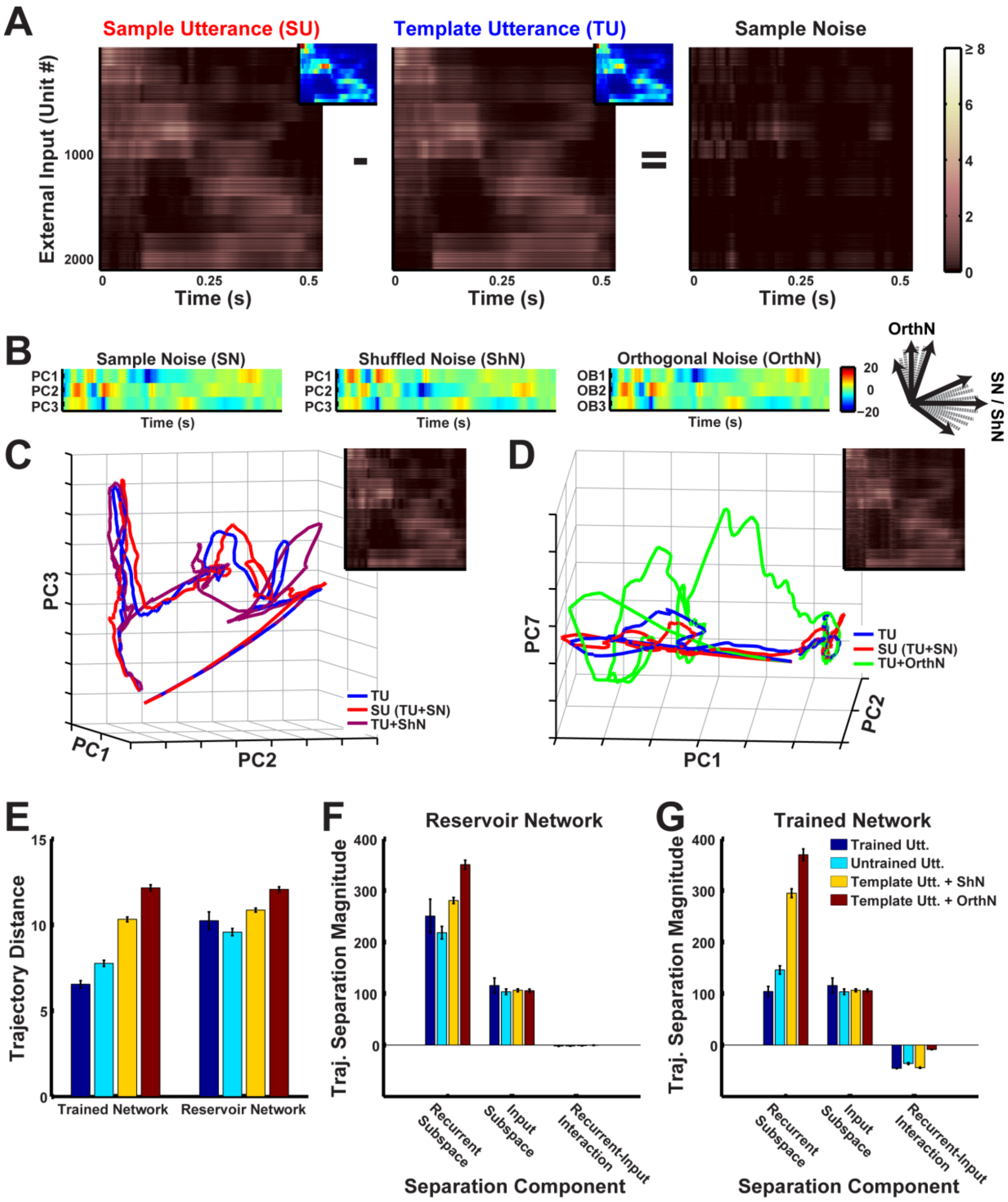

**Figure 6: Robustness to spectral noise depends on the spatiotemporal structure of the inputs that the network is exposed to during training.**

**(A)** Spectral noise in the inputs to an RNN (N=2100) during presentations of digit zero. Sample noise is the difference between the external input (each row reflects net external input to a unit in the RNN) for a sample and template utterance (corresponding cochleogram inputs are shown as insets). Absolute values of input projections are shown here for better visualization. **(B)** Coordinates of spectral noise in PCA space used to construct noisy utterances for digit zero. First three Principal component (PC) scores of the sample noise (left). Scores, where the basis was constructed as a random shuffle of the sample noise PC loadings (middle) and an orthonormal set orthogonal to the sample noise PC loadings (right). **(C-D)** Projections in PCA space of the external input for a template utterance, a sample utterance, and artificial utterances created from the shuffled basis noise (**C**) and orthogonal basis noise (**D**). The respective artificial external inputs are shown as insets. **(E)** Comparison of mean within-digit distances of the trajectories for the different natural (2 trained, blue and 7 untrained, cyan), and artificial utterances (30 shuffled, yellow and orthogonal, red each) utterances from the template utterance. Bars represent mean of the time-averaged distances for respective utterances of all digits (1 subject, 10 digits). Error bars indicate standard errors of the mean over the digits. **(F-G)** Magnitudes of components that contribute to the total within-digit distance measured in (E) in the reservoir **(F)** and trained **(G)** networks. Bars represent mean of the time-averaged values of the components for respective utterances of all digits (1 subject, 10 digits). Error bars indicate standard errors of the mean over the digits. The network was trained with $I_0$ set to 0.25, and trajectories were recorded with $I_0$ set to 0.

to multiple recurrent units via common input projections. Moreover, the spectral noise in different channels, particularly neighboring ones, exhibit structured time-dependent relationships; Third, spectral noise is proportional in magnitude to the input amplitude, and therefore tends to be much stronger than the background noise that is typically injected during training and testing.

To better characterize how spectral noise invariance emerges during training, we measured network responses while naturally and artificially varying the structure of the spectral noise. Any confounding effects from temporal invariance were avoided by training and testing the network on temporally normalized utterances, by warping them to the duration of the template utterance for the respective digit. Responses to standard natural test utterances (those not presented during training) were contrasted with two types of artificially created utterances: those with shuffled spectral noise and orthogonal spectral noise (**Fig 6B**). Shuffled spectral noise was created by shuffling the PCA axes (basis) of the spectral noise (Materials and Methods). In other words, novel utterances were generated by reordering the directions of spectral noise in phase space (**Fig. 6C**). This procedure scrambled the relationships between input signal and noise, while constraining the artificial spectral noise to the same subspace as natural spectral noise. Orthogonal spectral noise was composed of linear bases that were orthogonal to natural spectral noise (**Fig. 6D**, see Materials and Methods). In addition to scrambling the input signal-noise relationships, this procedure also scrambled the relationships between input channels. Care was taken while constructing these artificial inputs, to ensure that the temporal structure, dimensionality and magnitude of the spectral noise were identical to that observed in natural utterances.

Measurements of the within-digit trajectory distances revealed a progressive increase from trained utterances, to untrained utterances, to utterances with shuffled noise, to utterances with orthogonal noise (**Fig. 6E**), wherein the difference between the trained and untrained utterance conditions was small. In contrast, the untrained (reservoir) network exhibited deviations that were large and fairly similar across all conditions, implying that the trained network was able to identify and suppress natural spectral noise present in the utterances, but not artificial noise of similar magnitude and structure. We further dissected these results with a novel analysis based on the decomposition of the RRN drive into its input and recurrent components, which allowed trajectories to be decomposed into the subspaces wherein the network integrates external (input subspace) and recurrent (recurrent subspace) inputs (Materials and Methods). The analysis showed that in untrained networks, both natural and artificial spectral noise introduced deviations in the recurrent subspace that were large and unsuppressed (**Fig. 6F**). In contrast, in trained RNNs the recurrent dynamics suppressed natural spectral noise, but not artificial noise (**Fig. 6G**). Training also resulted in a recurrent subspace that suppressed spectral noise induced deviations in the input subspace. Whereas the untrained network integrated recurrent and external inputs in orthogonal subspaces, training the network rotated the recurrent subspace such that spectral noise induced trajectory deviations in the recurrent subspace were anti-correlated with the deviations in the input subspace, thereby resulting in their suppression (Recurrent-Input interaction in **Fig. 6F-G**).

Together, these results suggest that recurrent plasticity may play a crucial role in identifying and suppressing the natural spatial variations of stimuli. In our model, this is achieved via supervised training with a common within-digit target trajectory, which exposes the spatiotemporal distribution of the spectral noise in the sensory input, and in doing so, allows for effective sampling from this distribution. Training then shapes the basins of attraction or hypertubes around the spoken digit-encoding trajectories based on this distribution, resulting in effective, recurrently-driven suppression of naturally occurring spectral noise.

### *Encoding Stimuli as Neural Trajectories Allows for Temporal Generalization (Scaling).*

As mentioned above, a signature and little understood feature of how the brain processes time-varying patterns pertains to temporal warping (temporal invariance), whereby temporally compressed or dilated input signals can be identified as the same pattern. Thus an important test for computational models of sensory processing of time-varying stimuli, is whether they are able to account for temporal invariance. We hypothesized that one advantage of encoding time-varying stimuli as continuous neural trajectories, is that it naturally addresses the problem of temporal warping; specifically, that compressed or dilated stimuli generate similar neural trajectories that play out at different "speeds" (Lerner et al., 2014; Mello et al., 2015). To test this hypothesis we trained and tested an RNN on a dataset of temporally warped spoken digits. Specifically, the input pattern, $\boldsymbol{y}(t)$, of a single utterance of each digit from a speaker of the TI-46 dataset was artificially stretched or compressed by a time warp factor $\alpha$ ($\boldsymbol{y}(\frac{1}{\alpha}t)$), while retaining its spatial structure. **Figure 7A** (left panel) shows the input structure for two utterances of the digit “nine”, one warped to twice ($\alpha$=2, 2x or 200% warp) and the other to half ($\alpha$=0.5, 0.5x or 50% warp) the duration of the original utterance.

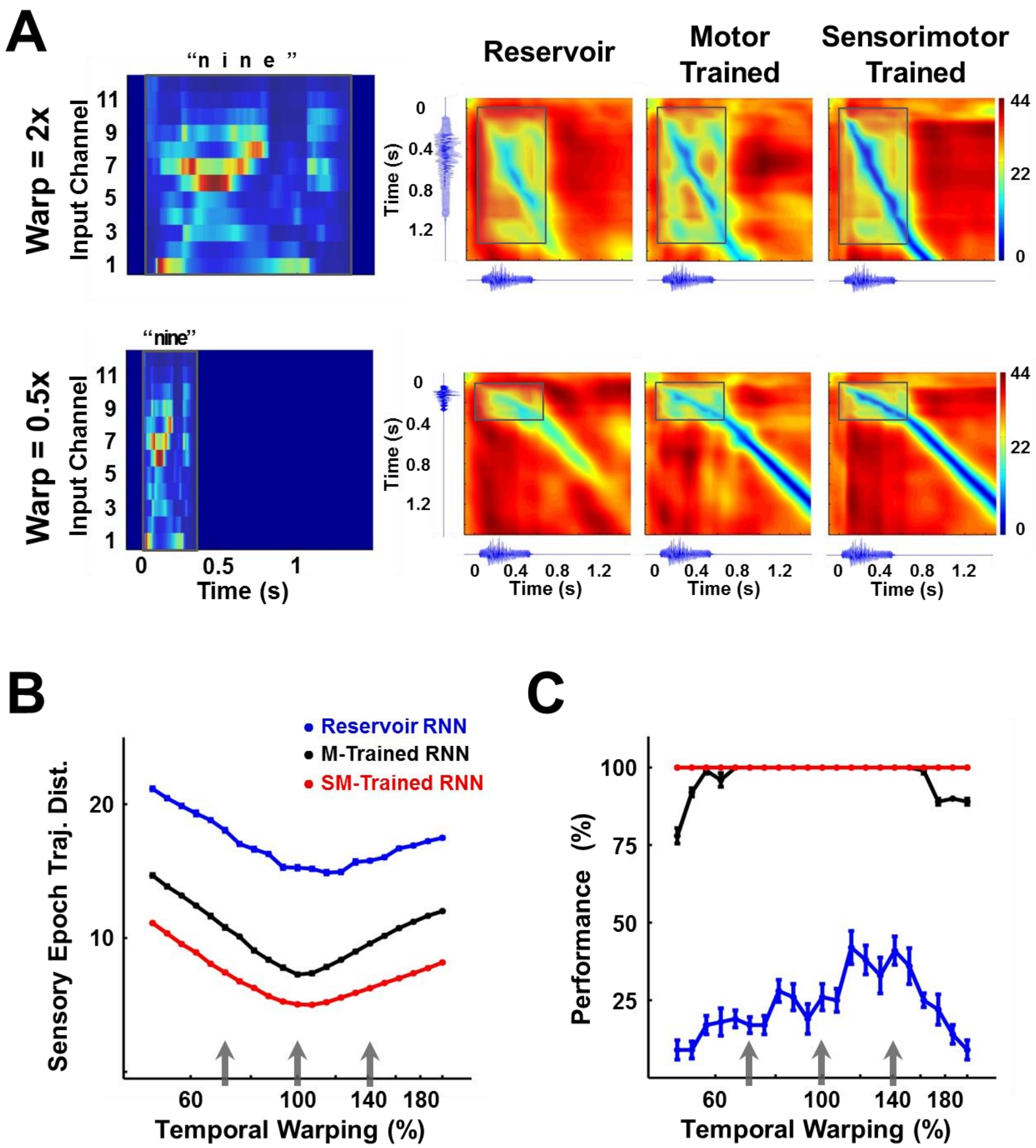

**Figure 7: Invariance of encoding trajectories to temporally warped spoken digits.**
(**A**) Temporally warped input cochleograms for an utterance of the digit "nine" (left panel), warped by a factor of 2x (upper row) and 0.5x (lower row). Distance matrices between the trajectories produced by the warped and reference stimuli for the two control networks and the sensorimotor trained network (right panels). Distance matrices are accompanied by the corresponding sonograms on each axis, and highlight the comparison between the sensory trajectories of the warped and reference utterance (outlined box). A deep blue diagonal through a distance matrix indicates that the sensory trajectory encoding the warped input overlaps with the reference trajectory, and thus that the trained RNN is invariant to temporal warping of the input. **(B)** Mean time-averaged Euclidean distance between sensory trajectories encoding warped and reference utterances of the 10 digits in each of the three networks, over a range of warp factors. Error bars (not visible) indicate standard errors of the mean over 10 test trials. Gray arrows indicate the warp factors at which the networks were trained. **(C)** Transcription performance (measured by the deep CNN classifier) of the three networks on utterances warped over a range of warp factors. The results corroborate the measurements in (B) and show that a sensorimotor trained RNN is more resistant to temporal warping. All networks were composed of 2100 units. Network training was performed with 30 utterances (1 subject, 10 digits 1 utterance per digit, 3 warps per utterance). $I_0$ was set to 0.25 during network training, and to 0.05 during output training and testing. Reference trajectories for the distance analysis were recorded with $I_0$ set to 0.

The RNN was trained on three utterances of each digit (warp factors of 0.7, 1 and 1.4) and tested on warp factors in the range 0.5 to 2 (a four-fold factor that approximates the natural speed variation of speech). To discern the effect of a temporally warped input on the encoding trajectory, we constructed a matrix of the distances between trajectories produced by the reference ($\alpha$=1) and a warped utterance (**Fig. 7A**). Each element of such a matrix measures the distance between the two trajectories at a corresponding pair of time points. In the case of perfect warping (e.g., when the reference trajectory and the trajectory at an $\alpha \neq 1$ overlap exactly in phase space), a straight diagonal line of zero distances (deep blue) would be observed during the sensory epoch (shaded region), with slope proportional to the warp factor. Trajectories produced by temporally warped inputs in the sensorimotor trained RNN were closer to the reference compared to their counterparts in the reservoir or motor-trained RNN. In other words tuning the recurrent weights dramatically improved the ability of the network to reproduce the same neural trajectories, despite being driven by inputs at different speeds. We quantified this effect by measuring the average distance between the reference and warped trajectories during the sensory epoch over different warp factors (**Fig. 7B**). In comparison to the two control networks, the trained RNN maintains a much smaller distance between trajectories across the four-fold range of temporal warping—including warp factors outside the range used during training. These results confirm the hypothesis that training produces a modulation of the internal dynamics by the external input that renders the encoding trajectories invariant to temporal warping of the input patterns. The impact of this result is borne out in the network's performance on the digit transcription task, as measured by the CNN classifier (**Fig. 7C**). Training during the sensory and motor epochs dramatically improves "extrapolation"—that is generalization to speeds outside the range of training speeds; however, training during the motor epoch alone is sufficient to produce very good "interpolation" (novel speeds within the training set range) generalization.

***Mechanisms Underlying Temporal Generalization (Temporal Scaling).***

It is surprising that a RNN with strong recurrent dynamics can represent temporally scaled stimuli with similar trajectories. Specifically, the strong intrinsic dynamics of the network (governed by the recurrent weights) must somehow match the temporal scale of the input. One possibility is that the network achieves this by scaling the linear speed of its trajectories (i.e. the magnitude of $\frac{dx}{dt}$ in Equation 1) in inverse proportion to the warp factor. However, a comparison of the time-averaged linear speed of trajectories encoding digits at different warps deviates from this relationship in both the untrained and trained networks (**Fig. 8A**), which, in contrast to our earlier results (**Figs. 4A, 7A**), suggests that the trajectories may not be temporally invariant. This contradiction arises from the incorrect assumption that the trajectories have little or no curvature. **Fig. 8B** schematizes the two alternatives for temporal scaling of trajectories with curvature: constant speed/variable distance traversed, versus variable speed/constant distance traversed. At one extreme, a reference trajectory (black curve) can be slowed down while conserving its linear speed by increasing its radius of curvature, and therefore the distance it traverses through phase space (green curve). At the other extreme, the reference may be slowed by appropriately scaling down its linear speed while conserving its radius of curvature (yellow curve), thereby generating temporally invariant trajectories that traverse identical distances.

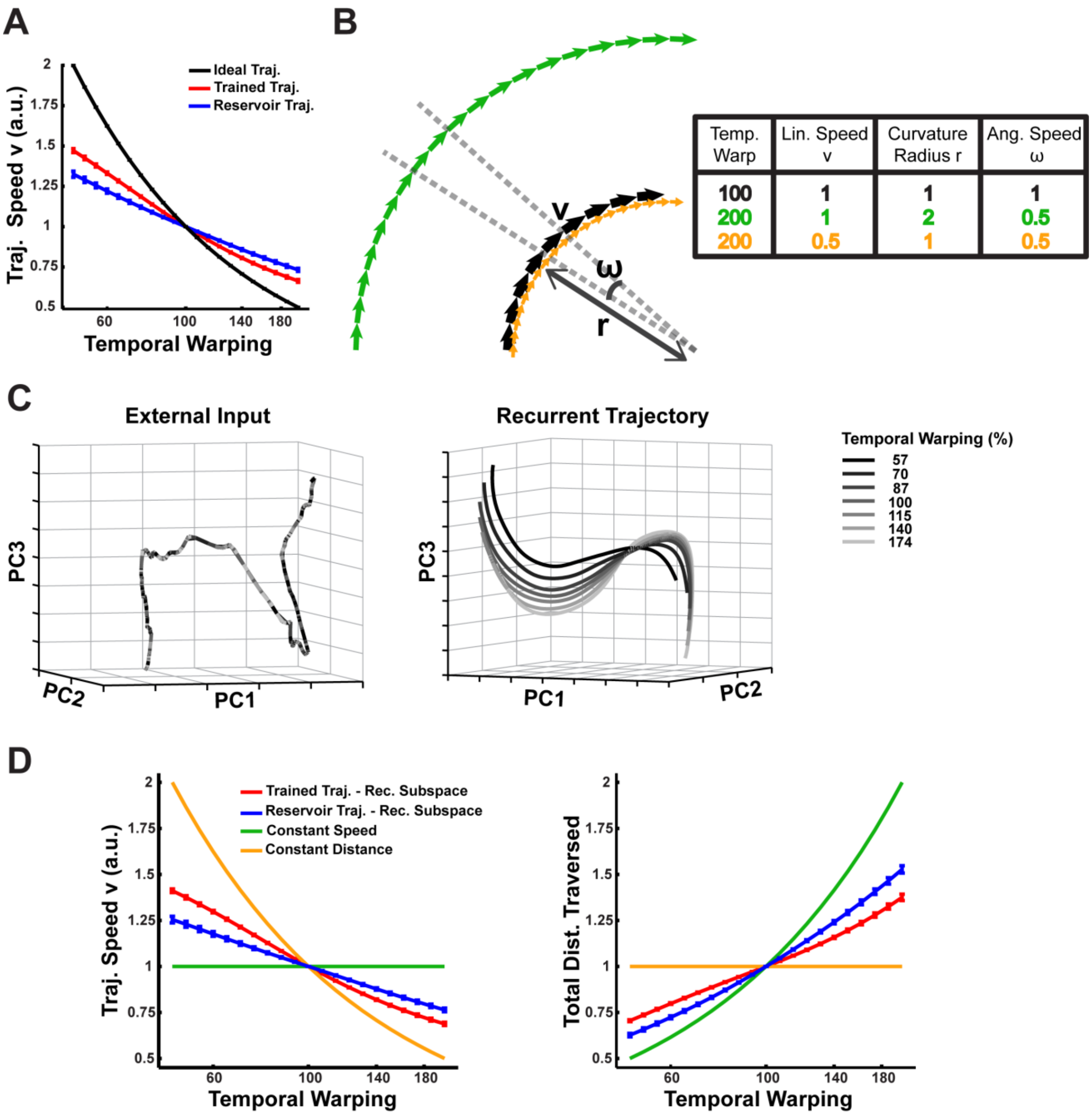

**Figure 8: Mechanism of temporal scaling invariance.**
**(A)** Time-averaged linear speed of the sensory epoch trajectories in the reservoir and trained networks (N=2100) compared to the ideal speed, over a range of warp factors. The speeds are normalized to the 1x warp. **(B)** Schematic representation of two hypotheses for trajectories that exhibit temporal warp invariance: constant speed/variable distance (green) and variable speed/constant distance (yellow) **(C)** Projections in PCA space of external input for warped utterances of digit one (left) and the corresponding sensory epoch trajectories of the trained network (right), over a 100ms duration of the 1x warp trajectory and the corresponding utterance interval of the remaining trajectories. Over this short duration, over 90% of the variance in the external input and the trajectories was captured. **(D)** Mean linear speed (left) and cumulative distance traversed (time integral of the linear speed; right) in the recurrent subspace of phase space of a reservoir and trained network, compared to predictions from the constant speed and constant distance hypotheses. The measures are normalized to the 1x warp. In all panels error bars indicate SEM over the digits. The network was trained with $I_0$ set to 0.25, and trajectories were recorded with $I_0$ set to 0.

As expected in PCA subspace, the temporally-scaled utterances produce overlapping trajectories—reflecting the scaling of their linear velocities while traversing constant distances (a direct consequence of how the artificially warped utterances were constructed). In contrast, a similar visualization of the respective trajectories of the trained network showed that they are spatially distinct with constant warp-dependent separations (**Fig. 8C**). Moreover, they exhibit warp-dependent curvature, with slower trajectories incurring larger curvature radii. Together with the results in **Fig. 8A**, this suggests that the mechanisms underlying temporal scaling in our model falls between the two extreme hypotheses, implying that: (i) that the average linear speed and curvature radius of trajectories encoding utterances at different warps adjust in a manner that appropriately modulates their angular speed and therefore their timescale; and (ii) that the trajectories are parallel to each other, demonstrating that they encode the digit similarly. **Figure 8D** presents measurements of the time-averaged linear speed and the total distance traversed in the recurrent subspace of phase space, by the trajectories in the trained and untrained networks over a range of warps. The latter is a heuristic measure of the time-averaged curvature radius, hinging on the fact that the circumference (or distance traversed) linearly scales with the radius. A comparison of these measurements for the two networks to the expected values in constant speed- and constant distance trajectories, confirms that both networks produce trajectories that lie in between these two extremes. That is, temporal scaling is achieved by an approximate balance of both hypotheses, e.g., at low speeds (digits spoken slowly) the trajectories are slower and longer. Furthermore, training alters this balance to reduce the within-digit separation, by modulating the trajectory speeds more strongly. Measurements to establish whether the trajectories are parallel also agree with these results (**Supplementary Fig. 3C**).

But how do identical inputs (except for their duration) generate spatially distinct trajectories (e.g., **Fig. 8C**)? To answer this question we first measured the relationship between the input and recurrent subspaces in the trained network, and discovered that they are orthogonal to one another (data not shown), implying that the integration of the external inputs is independent of the integration of the recurrent inputs. In the absence of subspace interactions, it is easy to see that external inputs with different speeds are integrated into spatially distinct input subspace trajectories (e.g. $\int \cos(\omega t)dt = \frac{1}{\omega}\sin(\omega t) + c$, wherein high speed/frequency signals integrate to produce trajectories with small magnitudes) (**Supplementary Fig.3A**). As with the recurrent subspace trajectories, the resulting input subspace trajectories are parallel to each other and modulate their angular speed via a combination of changes to their curvature radii and linear speed (**Supplementary Fig.3A-C**). Ultimately, the differentiated curvature and linear velocity that modulate the angular velocity and therefore the timescales of recurrent subspace trajectories, are directly shaped by these input subspace dynamics (Rajan et al., 2010b).

In summary, these results demonstrate that the integration of inputs at different warps separates out the resulting trajectories in the input subspace, and consequently in the recurrent subspace. In both subspaces the network dynamics generate trajectories with warp-dependent linear velocities and curvature, which appropriately modulate their angular velocity, resulting in them traversing spatially distinct yet parallel paths at warp-dependent timescales. This confirms the hypothesis that an inherent computational advantage of encoding time-varying sensory stimuli in neural trajectories is that it can naturally address the temporal invariance problem.

**Discussion:**

The brain naturally encodes, recognizes, and generates complex time-varying patterns, and can seamlessly process temporally scaled patterns. Despite our poor understanding of these processes, theoretical evidence increasingly suggests that the dynamics of recurrently connected circuits in the brain are critical to representing time-varying patterns. For example, so-called reservoir computing approaches propose that complex high-dimensional spatiotemporal patterns are represented in the dynamics inherent to randomly connected recurrent neural networks (Jaeger and Haas, 2004; Buonomano and Maass, 2009; Sussillo and Abbott, 2009). A shortcoming of this approach, however, is that the recurrent connections in these networks are not plastic; thus in contrast to actual cortical circuits, the random recurrent neural network (the "reservoir") does not adapt or optimize to the task at hand. Additionally, fully tapping into the computational potential of randomly connected RNNs has proven difficult because they are susceptible to chaos (Sompolinsky et al., 1988b; Wallace et al., 2013). However, progress has been made on both accounts (Martens and Sutskever, 2011; Vogels et al., 2011; Laje and Buonomano, 2013). Here, we extend these results to better understand well-known aspects of the brain's ability to represent time-varying sensory *and* motor patterns—we demonstrate how experience-dependent plasticity could reshape the dynamics of reservoir RNNs to qualitatively improve representations by endowing them with such essential properties as generalization to novel exemplars, sensory-motor association and transformation, and temporal invariance.

Recent studies have enlisted synaptic dynamics to achieve temporal warping (Gütig and Sompolinsky, 2009; Murray and Escola, 2017). In contrast, the model described here illustrates that encoding temporal patterns within the neural trajectories of recurrent networks naturally enables dynamic pattern recognition of time-warped stimuli. Specifically, the integration of temporally scaled stimuli yields proximal parallel trajectories with scale-dependent angular velocities. The formation of such trajectories is surprising given the sensitivity and nonlinear nature of RNNs (as shown by the performance of the reservoir network). Yet, by training the RNN, it is possible to reshape its dynamics such that temporally warped sensory stimuli are encoded by similar trajectories in neural phase space that exhibit scale-dependent angular velocities.

Our results also demonstrate how training an RNN reveals a computationally powerful input processing regime, one that helps generalize the encoding of learned time-varying patterns to novel exemplars. As stated earlier, recognizing different instances of the same digit as one and the same, while discriminating between different digits requires that networks actively and differentially sculpt the within and between class trajectories—the separation between trajectories representing similar patterns must be kept to a minimum, while that between trajectories representing different patterns must be amplified. A previous study has analytically shown that untrained random recurrent networks (reservoir networks) process inputs in one of three computational regimes: a chaotic regime, wherein any separation between trajectories is strongly amplified by the network's chaotic dynamics; a regime with weak recurrent weights that is input-dominated with little computational power; and a critical regime wherein the input and internal dynamics are balanced (Bertschinger and Natschläger, 2004). Here, we show that training alters these regimes in very different ways (**Fig. 5**). Crucially, relative to the separation between the input patterns, reservoir networks in the critical regime amplify the separation between trajectories representing similar patterns more strongly than between different patterns. In contrast, trained networks separate trajectories according to the patterns they are representing (**Fig. 5A**). Tuning the recurrent weights reshapes the internal dynamics of the network, enabling it to redirect or suppress deviations from the target trajectories (**Fig. 6**). Ultimately, it is this property of trained networks that allows them to encode complex sensory patterns and effectively generalize across natural utterances and speakers. Conversely, simulations with artificially generated inputs demonstrate that trained networks fail at generalizing to unfamiliar spectral noise patterns, predicting similarly divergent trajectories in sensory areas when presented with artificial stimuli of this nature.

The computational potential of continuous time RNNs in high-gain regimes has long been recognized, but it has been challenging to tap into this potential because of their inherently chaotic behavior and the difficulties in training them (Bengio et al., 1994; Pearlmutter, 1995). Step-by-step, progress has been made on how to capture the computational potential of RNNs (Jaeger and Haas, 2004; Sussillo and Abbott, 2009; Martens and Sutskever, 2011; Laje and Buonomano, 2013). Here we establish that a further feature of trained RNNs is their ability to not only encode spatiotemporal objects, but to perform complex sensorimotor tasks and address the long-standing problem of temporal warping. We propose that spatiotemporal objects are not encoded as fixed-point attractors, but as locally stable neural trajectories (dynamic attractors). Two experimentally tractable predictions emerge from our results. First, neural trajectories will be stable in response to local perturbations—potentially administered through optogenetic stimulation—and that the neural trajectories elicited during sensory stimuli will be more resistant to perturbations than the neural trajectories unfolding during the motor epoch of sensory-motor tasks. Second, the model predicts a clear neural correlate of the encoding of temporally scaled stimuli. Specifically, the observation that slower stimuli yield trajectories with larger curvature radii implies that the neural population activity should exhibit larger fluctuations in their firing rates in response to slower stimuli, in other words, at the neuronal level the range of minimal to maximal firing rate should be larger (**Supplementary Fig. 3D**).

**Materials and Methods:**

Network Model

The dynamics of the RNN was comprised of *N* nonlinear continuous-time firing rate units modeled as:

$$\tau \frac{dx_i}{dt} = -x_i + \sum_{j=1}^{N} W_{ij}^{R} r_j + \sum_{j=1}^{M} W_{ij}^{I} y_j + I_i^{noise} \qquad (1)$$

$$r_i = \tanh(x_i) \qquad (2)$$

$x_i$ represents the state of neuron *i*, and $r_i$ its "firing rate". The time constant, $\tau$, of each unit was set to 25 ms. $\boldsymbol{W^R}$ is the weight matrix representing the recurrent connectivity of the network. The recurrent connectivity was uniformly random (but with no autapses) with a connection probability ($p_c$) of 0.2. The weights of these synapses were initialized from an independent Gaussian distribution with zero mean and SD equal to $g/\sqrt{p_C N}$ , where *g* represents the "gain" of the network. RNNs were initialized to a high-gain regime (*g* = 1.6), which generates chaotic self-perpetuating activity in the absence of external input or noise (Sompolinsky et al., 1988b).

The M-dimensional vector ***y(t)*** represents the time-varying sensory input to the RNN. The fixed input weight matrix, $\boldsymbol{W^I}$, tonotopically projects this input vector onto the RNN—input channel *k* is projected onto units (k-1)×N/M+1 thru k×N/M. The weights of these connections were drawn from an independent Gaussian distribution with zero mean and unit variance. $\boldsymbol{W^I}$ ***y(t)*** represents the net external input to each RNN. The handwriting was modeled with 3 output units ($o_1$, $o_2$, $o_3$) that represented the x, y and z co-ordinates of a pen on paper. Finally, each unit of the RNN also received an independent background noise current, $\boldsymbol{I^{noise}}$, modeled as additive Gaussian white noise with SD $I_0$.

As is standard, the output neurons were simulated as a weighted linear sum of all the units in the RNN:

$$o_i = \sum_{j=1}^{N} W_{ij}^{O} r_j \qquad (3)$$

where the output weight matrix, $\boldsymbol{W^O}$, was initialized from an independent Gaussian distribution with zero mean and SD $1/\sqrt{N}$ . All simulations were performed with a time step of 1 ms.

Simulations and Training

Each trial consisted of a time window comprised of a sensory and a motor epoch. We defined the sensory epoch as the period in which an external stimulus is presented—starting at *t*=0 and lasting the duration of the utterance. The motor epoch was defined as a period beginning 300 ms after the sensory epoch ended and lasting the duration of the target motor pattern (the appropriate handwritten digit).

Training proceeded in three steps (described in detail below): (1) target trajectories were generated for each utterance of each digit in the training subset of the dataset; (2) the recurrent units were trained to reproduce the target trajectories; and (3) the output units were trained to produce the handwritten spatiotemporal patterns. The trained network was then tested on novel (and trained) utterances. In steps 2, 3 and during testing, each trial began at $t$=-100 ms with the network initialized to a random state ($x_i$ values drawn from a uniform distribution between -1 and 1). In **Figs. 1**, **7** and **Supplementary Fig. 1**, testing was performed with the noise amplitude ($I_0$) set to 0.05. For the perturbation analysis (**Fig. 2**), a 25 ms "perturbation pulse" was introduced during the sensory or motor epoch of each trial, with $I_0$ set to the desired perturbation magnitude for the duration of the pulse. Noise was omitted in these simulations to allow for a direct assessment of the effects of the perturbation pulse. Similarly, for the simulations in **Figs. 3-6, Fig. 8** and **Supplementary Fig. 3**, noise was omitted ($I_0$=0) so that the impact of training on the generalization and discriminability of an RNN's encodings, but not its background noise invariance, could be direct evaluated.

Simulations of the untrained "reservoir" control network skipped steps 1 and 2, while simulations of the motor-trained control network limited recurrent network training (step 2) to a duration starting 150 ms after the end of the sensory epoch and lasting until the end of the motor epoch.

Input Structure

We used spoken digits from the TI-46 spoken word corpus (Mark Liberman, 1993) to train and test networks on the transcription task. Specifically, our dataset was composed of the spoken digits "zero" thru "nine", uttered 10 times each, by each of five female subjects. Spoken digits from the corpus were decoded, end-pointed, resampled to 12 kHz, and converted to spectrograms with Matlab's specgram function. The spectrograms were preprocessed with Lyon's passive ear model of the human cochlea (Lyon, 1982), as implemented by the auditory toolbox (Malcolm Slaney), to generate analog "cochleograms" composed of 12 analog frequency bands, or channels, ranging from 0 to 6 kHz. Finally, the cochleograms were smoothed with a 20$^{th}$ order 1D median filter (Matlab's medfilt1 function), normalized to a maximum of 1 (i.e., they were normalized by the maximal value of all utterances and digits), and scaled by an input amplitude. During the sensory epoch of each trial, the input, $\boldsymbol{y(t)}$ (M=12), took values from the cochleogram corresponding to the utterance that was to be presented to the RNN. At all other times of the trial, $\boldsymbol{y(t)}$ was set to 0. The input amplitude was set to 5 in all simulations except in **Fig. 5**, where it was parametrically varied. For the temporal warping simulations (**Figs. 7-8**) the cochleograms were compressed or dilated by the warping factor $\alpha$ through linear interpolation.

Innate Trajectories and RNN Training

Training was performed with the "innate-learning" approach—which uses the Recursive Least Squares (RLS) rule to train each unit of the recurrent network to match the pattern generated by the untrained network (Laje and Buonomano, 2013). For each subject used in the training set, a single template utterance of each digit was presented to the untrained RNN (the reservoir) in the absence of background noise, and the resulting trajectory served as a target ("innate") trajectory. The template utterance of each digit was chosen as one with the median duration among all the utterances of the digit by the subject. Target sensory epoch trajectories for other training utterances of the digit by the same subject were generated by linearly warping this innate trajectory to match the utterance durations.

The initial network state was chosen at random while harvesting the target trajectory for each digit. Target trajectories for template utterances of the same digit by different speakers were harvested starting the network at the same initial state. To achieve the formation of digit-specific dynamic attractors, the same target motor trajectories were used for all utterances of each digit. All networks were trained with three utterances of each digit for each of the subjects in the training set. The networks in **Fig. 1** and **Supplementary Fig. 1** were trained on three subjects and tested on five, while those in **Figs. 2-8** and **Supplementary Fig. 3** were trained and tested on 1 subject.

Network training was performed by modifying the recurrent weight matrix, $\boldsymbol{W^R}$, with the Recursive Least Squares learning rule (Haykin, 2002). The rule was simultaneously applied to 90% of the units in the network (randomly selected). Training was conducted by iterating through all utterances of the digits in the training set over multiple trials, starting each trial at a random initial condition and continuously injecting the network with background noise ($\boldsymbol{I^{noise}}$). Training concluded when the error in the activity of the rate units asymptoted (generally between 100 and 150 trials).

<u>Output Training</u>

The spatiotemporal target patterns that comprise the handwritten digits were sampled from a Wacom Bamboo Pen Tablet, as the digits “0” thru “9” were individually stenciled on it. For each handwritten digit, the x and y coordinates of the pen were sampled at approximately 50 Hz, low-pass filtered, and resampled with interpolation to 1 kHz (corresponding to the 1 ms simulation time step). The target values for $o_1$ and $o_2$ were set to 0 from the start of each trial until the beginning of the motor epoch. During the motor epoch, they were set to the pen’s 2D coordinates for the corresponding digit, and reset to 0 between the end of the motor epoch and the end of the trial. The target for $o_3$ (z co-ordinate) was a step function, set to 1 during the motor epoch and 0 at all other times. To train the output units, the Recursive Least Squares learning rule was applied to the readout weights in $\boldsymbol{W^O}$ (Haykin, 2002; Jaeger and Haas, 2004; Sussillo and Abbott, 2009; Laje and Buonomano, 2013). Output training was performed for 25 trials per utterance of each digit in the training set, while the RNN was continuously injected with background noise.

In each test trial, the motor output was recorded from the values of $o_1$ and $o_2$, whenever the value of $o_3$ was greater than 0.5 (i.e. when “the pen contacted the paper”). At the end of the trial, a 28×28 pixel grayscale image of the “handwritten” output (pen width = 2 pixels) was labeled as the transcription for the corresponding digit. An objective determination of the transcription’s accuracy was made by comparing this label to one assigned to the image by a CNN classifier for handwritten digits. This classification was performed with the LeNet-5 CNN classifier (Lecun et al., 1998a) implemented with the Caffe deep learning framework (Jia, 2013). The CNN was first trained on the MNIST database of handwritten digits (LeCun et al., 1998b), and its output layer was then fine-tuned on the stenciled digits that we used as the output targets.

Trajectory Analysis

In **Figs. 4-5**, the sensory (motor) epoch within-digit distances were calculated from the Euclidean distance, at each time step, between the sensory (motor) epoch trajectory for each utterance, and its nearest neighbor from among the trajectories produced by the training utterances of the same digit. Similarly, the sensory (motor) epoch between-digit distances were calculated from the Euclidean distance, at each time step, between the sensory (motor) epoch trajectory for each utterance, and its nearest neighbor from among the sensory (motor) epoch trajectories encoding training utterances of all other digits. Similarly, in **Fig. 6E**, sensory epoch within-digit distances were calculated as the Euclidean distance between the sensory epoch trajectory for each tested external input, and the one encoding the template utterance. Finally, trajectory distances in **Fig. 7B** were also calculated in a similar fashion: at each time step, the Euclidean distance was calculated between the trajectory encoding an utterance at warp factor $\alpha$ ($I_0$=0.05), and its nearest neighbor along the trajectory encoding the reference utterance of the same digit ($I_0$=0). The time-average of these distances were then summarized over 10 trials for the plot in **Fig. 7B**.

The nature and robustness of spectral generalization in trained RNNs was probed in **Fig. 6**, by artificially altering the external inputs. In order to assess spectral generalization independent of temporal invariance, for each digit, the duration of all its utterances were time-normalized by warping the respective cochleograms to their median duration. The RNN and its outputs were trained on three time-normalized utterances for each of the ten digits, and all test simulations were performed with a common initial state. The spectral generalization of an RNN was tested with artificially constructed external inputs to the network that were qualitatively similar to external inputs for time-normalized natural utterances, but with altered spectral noise structure. The spectral noise in an utterance was defined as the difference between the external inputs of the utterance and the respective template utterance (**Fig. 6A**). Network responses to time-normalized trained and untrained natural utterances were compared to two artificial controls: external inputs with shuffled noise and with orthogonal noise. Inputs with shuffled noise were constructed from the Principal Component Analysis (PCA) loadings and scores of the spectral noise in untrained natural utterances as follows: (i) the loading vectors were permuted; (ii) shuffled noise was generated by multiplying the spectral noise scores by the shuffled loading vectors; (iii) this shuffled noise was added to the external input of the corresponding template utterance. Inputs with orthogonal noise were also constructed from the PCA loadings and scores of the spectral noise in untrained natural utterances, except instead of permuting the loading vectors, they were replaced by an orthonormal set of vectors, generated via Gram-Schmidt orthogonalization, each of which was orthogonal to the set of spectral noise PCA loading vectors. For this a random vector was generated and QR factorization was performed on the set of vectors comprised of the PCA loadings and the random vector, producing a new vector that was orthogonal to the PCA loadings. This procedure was repeated to generate a set of such vectors equal in cardinality to the PCA loading set. The random vectors were then orthogonalized relative to each other in a similar manner, and then normalized to produce an orthonormal set. External inputs with shuffled and orthogonal noise were constructed from each untrained natural utterance, with as few PCs as were necessary to explain 99% of the variance in its spectral noise (typically 11 PCs). For each digit, the network was tested with 30 shuffled and orthogonal noise inputs, each based on a randomly chosen untrained natural utterance.

The dimensionality measure shown in **Fig. 5** was calculated as $\left(\sum_{k=1}^{N}\lambda_k^2\right)^{-1}$, where $\lambda_k$ represents eigenvalues of the equal-time cross-correlation matrix of network activity, expressed as a fraction of their sum (Rajan et al., 2010a). The eigenvalues were calculated on a concatenation of the sensory epoch trajectories for all utterances (trained and novel) of all digits.

RNN Decomposition

In **Figs. 6 and 8**, trajectories and their distances from each other were decomposed into the constituent input and recurrent subspaces of phase space. These are derived from the state variable ***x(t)*** (Equation 1), rather than the rate variable ***r(t)*** (Equation 2) that was used to measure Euclidean distances:

From Euler's Method:

$$x(t) = x(t-1) + \frac{dx(t-1)}{dt}$$

$$x(t) = \frac{\tau-1}{\tau}x(t-1) + \frac{1}{\tau}W^R r(t-1) + W^I y(t-1)$$

Let $a = \frac{\tau-1}{\tau}$, $b = \frac{1}{\tau}$, $R(t) = W^R r(t-1)$, and $I(t) = W^I y(t-1)$

$$x(t) = ax(t-1) + bR(t) + bI(t)$$

Solving the recurrence relationship,

$$x(t) = a^t x(0) + b\sum_{k=1}^{t} a^{t-k} R(k) + b\sum_{k=1}^{t} a^{t-k} I(k) \quad (4)$$

where $x(0)$ is the initial state of the network. We denote the first two terms ($a^t x(0) + b\sum_{k=1}^{t} a^{t-k} R(k)$) as network activity in the recurrent subspace (recurrent subspace trajectory) and the last term ($b\sum_{k=1}^{t} a^{t-k} I(k)$) as network activity in the input subspace (input subspace trajectory).

When the input durations for all utterances of a digit are time-normalized, as in **Fig. 6**, the squared Euclidean distance between the state variables $x^1(t)$ and $x^2(t)$ for two sensory epoch trajectories encoding utterances $y^1(t)$ and $y^2(t)$ is given by:

$$\Delta x(t).\Delta x(t) = (x^2(t) - x^1(t)).(x^2(t) - x^1(t))$$

Let $\overline{\Delta R(t)} = b\sum_{k=1}^{t} a^{t-k}(R^2(k) - R^1(k))$ and $\overline{\Delta I(t)} = b\sum_{k=1}^{t} a^{t-k}\left(I^2(k) - I^1(k)\right)$ (**Supplementary Fig. 2**). Then assuming common initial state,

$$\Delta x(t).\Delta x(t) = (\overline{\Delta R(t)} + \overline{\Delta I(t)}).(\overline{\Delta R(t)} + \overline{\Delta I(t)})$$

$$\Delta x(t).\Delta x(t) = \overline{\Delta R(t)}.\overline{\Delta R(t)} + \overline{\Delta I(t)}.\overline{\Delta I(t)} + 2\overline{\Delta R(t)}.\overline{\Delta I(t)} \quad (5)$$

where the first ($\overline{\Delta R(t)}.\overline{\Delta R(t)}$) and second ($\overline{\Delta I(t)}.\overline{\Delta I(t)}$) terms denote the squared distance in recurrent subspace and input subspace, respectively. The third term ($2\overline{\Delta R(t)}.\overline{\Delta I(t)}$) is based on the interaction between the two subspaces: If the recurrent and input subspaces are orthogonal to each other, then this term will be zero; otherwise, it indicates whether the deviations in the recurrent subspace serve to amplify or suppress the spectral noise.

In **Fig. 8** and **Supplementary Fig. 3**, all test simulations were performed with a common initial state. The linear speed of the trajectory in neural phase space, and in its input and recurrent subspaces, were calculated as the time-averaged magnitude, or $L_2$-norm, of the instantaneous change in network state ($\frac{dx}{dt} = x(t+1) - x(t)$), and its input and recurrent subspaces projections, respectively (Equation 4). Similarly, the total distance traversed in each subspace was calculated as the sum of the magnitude of instantaneous change in network state in the respective subspace, over the duration of the warped utterance.

Analysis of Parallel Trajectories

In **Supplementary Fig. 3**, we present evidence in support of parallel trajectories produced by the network dynamics in response to warped utterances. For each digit, the trajectories were first temporally aligned to the 100% (1x) warp, by matching the time indices of the respective warped utterances to the 100% template. The following procedure was then independently applied to the temporally aligned recurrent and input subspace trajectories. One separation vector, $\mathbf{S}_{0.7x}(t_{aligned})$ ($\mathbf{S}_{1.4x}(t_{aligned})$), was calculated at each aligned time point of the trajectory at the fast 0.7x (slow 1.4x) warp, as the difference between the population state ($\boldsymbol{x}$) of the 0.7x (1.4x) warp and 1x warp trajectories. The separation vectors were then normalized to unit length, denoted as $\mathbf{S}_{0.7x}^{norm}(t_{aligned})$ ($\mathbf{S}_{1.4x}^{norm}(t_{aligned})$). Finally, the projected separations at each fast (slow) warp were generated by calculating the respective separation vectors and projecting them onto the normalized separation vectors of the 0.7x (1.4x) warp trajectory. For example, projected separations at the 0.5x warp, $PS_{0.5x}(t_{aligned})$, were generated by projecting the separation vectors for the 0.5x warp trajectory, onto $\mathbf{S}_{0.7x}^{norm}(t_{aligned})$, to give $PS_{0.5x}(t_{aligned}) = \mathbf{S}_{0.5x}(t_{aligned}).\mathbf{S}_{0.7x}^{norm}(t_{aligned})$. The variance explained by these projections was calculated as the ratio between the total population variance of the projected and overall separation.

Trajectory Visualization

For **Fig. 3**, PCA was performed on a concatenation of the trajectories generated by both the reservoir and the trained network in response to all utterances of the digits “six” and “eight” by a single subject. Trajectories were then individually projected onto three principal components (PCs) and plotted in 3D. The sensory (motor) trajectories were projected onto PCs 2-4 (PCs 1-3). PC 1 was not used in plotting the sensory trajectories, because it captured features common to both spoken digits. Similarly, in **Fig. 6**, PCA was performed on a concatenation of the external inputs for the template utterance of digit zero, a natural test utterance of digit zero by the same subject, and one artificial test utterance each (shuffled and orthogonal noise) derived from these utterances. The input projections were then projected on the first three PCs at plotted in 3D. In **Fig. 8**, PCA was performed on concatenations of temporally-aligned 100ms segments (i.e. segments aligned to 200-300ms of the 1x warp) of the external input for digit one at warps of 0.57x, 0.7x 0.87x, 1x, 1.15x, 1.4x and 1.74x. The external input segments where then projected onto the first three PCs and plotted in 3D. The same procedure was followed in plotting the PCA projections of population state responses in **Fig. 8,** and of the recurrent and input subspace population state in **Supplementary Fig. 3**.

**Acknowledgements:**

This research was supported by the NSF (IIS-1420897) and a Google Faculty Research Award. We thank Nicholas Hardy, Alexandre Rivkind and Jonathan Kadmon for helpful discussions, and Dharshan Kumaran for comments on an earlier version of this manuscript.

**Competing Financial Interests:**

None.

## References:

Abbott, L.F., DePasquale, B., and Memmesheimer, R.-M. (2016). Building functional networks of spiking model neurons. Nat Neurosci 19, 350-355.

Amit, D.J., and Brunel, N. (1997). Model of global spontaneous activity and local structured activity during delay periods in the cerebral cortex. Cereb Cortex 7, 237-252.

Ayaz, A., Saleem, Aman B., Schölvinck, Marieke L., and Carandini, M. (2013). Locomotion Controls Spatial Integration in Mouse Visual Cortex. Current Biology 23, 890-894.

Bengio, Y., Simard, P., and Frasconi, P. (1994). Learning long-term dependencies with gradient descent is difficult. IEEE Trans Neural Netw 5, 157-166.

Bertschinger, N., and Natschläger, T. (2004). Real-time computation at the edge of chaos in recurrent neural networks. Neural Comput 16, 1413-1436.

Buonomano, D.V., and Maass, W. (2009). State-dependent Computations: Spatiotemporal Processing in Cortical Networks. Nat Rev Neurosci 10, 113-125.

Buonomano, D.V., and Merzenich, M.M. (1998). Cortical plasticity: from synapses to maps. Annual Rev Neuroscience 21, 149-186.

Carnevale, F., de Lafuente, V., Romo, R., Barak, O., and Parga, N. (2015). Dynamic Control of Response Criterion in Premotor Cortex during Perceptual Detection under Temporal Uncertainty. Neuron 86, 1067-1077.

Chang, E.F., Niziolek, C.A., Knight, R.T., Nagarajan, S.S., and Houde, J.F. (2013). Human cortical sensorimotor network underlying feedback control of vocal pitch. Proceedings of the National Academy of Sciences 110, 2653-2658.

Cheung, C., Hamiton, L.S., Johnson, K., and Chang, E.F. (2016). The auditory representation of speech sounds in human motor cortex. eLife 5, e12577.

Crist, R.E., Li, W., and Gilbert, C.D. (2001). Learning to see: experience and attention in primary visual cortex. Nature neuroscience 4, 519-525.

Crowe, D.A., Zarco, W., Bartolo, R., and Merchant, H. (2014). Dynamic Representation of the Temporal and Sequential Structure of Rhythmic Movements in the Primate Medial Premotor Cortex. The Journal of Neuroscience 34, 11972-11983.

Doupe, A.J., and Kuhl, P.K. (1999). Birdsong and human speech: common themes and mechanisms. Annu Rev Neurosci 22, 567-631.

Elman, J.L. (1990). Finding Structure in Time. Cog Sci 14, 179-211.

Feldman, D.E., and Brecht, M. (2005). Map plasticity in somatosensory cortex. Science 310, 810-815.

Gütig, R., and Sompolinsky, H. (2009). Time-Warp–Invariant Neuronal Processing. PLoS Biology 7, e1000141.

Haykin, S. (2002). Adaptive Filter Theory (Upper Saddle River: Prentice Hall).

Hinton, G., Li, D., Dong, Y., Dahl, G.E., Mohamed, A., Jaitly, N., Senior, A., Vanhoucke, V., Nguyen, P., Sainath, T.N., and Kingsbury, B. (2012). Deep Neural Networks for Acoustic Modeling in Speech Recognition: The Shared Views of Four Research Groups. Signal Processing Magazine, IEEE 29, 82-97.

Hopfield, J.J. (1982). Neural networks and physical systems with emergent collective computational abilities. Proc Natl Acad Sci U S A 79, 2554-2558.

Hopfield, J.J. (2015). Understanding Emergent Dynamics: Using a Collective Activity Coordinate of a Neural Network to Recognize Time-Varying Patterns. Neural Computation 27, 2011-2038.

Hopfield, J.J., and Tank, D.W. (1986). Computing with neural circuits: a model. Science 233, 625-633.

Ivry, R.B., and Schlerf, J.E. (2008). Dedicated and intrinsic models of time perception. Trends in Cognitive Sciences 12, 273-280.

Jaeger, H., and Haas, H. (2004). Harnessing nonlinearity: predicting chaotic systems and saving energy in wireless communication. Science 304, 78-80.

Jia, Y. (2013). Caffe: An open source convolutional architecture for fast feature embedding. p. Caffe: An open source convolutional architecture for fast feature embedding.

Karmarkar, U.R., and Dan, Y. (2006). Experience-dependent plasticity in adult visual cortex. Neuron 52, 577-585.

Laje, R., and Buonomano, D.V. (2013). Robust timing and motor patterns by taming chaos in recurrent neural networks. Nat Neurosci 16, 925-933.

LeCun, Y., Bengio, Y., and Hinton, G. (2015). Deep learning. Nature 521, 436-444.

Lecun, Y., Bottou, L., Bengio, Y., and Haffner, P. (1998a). Gradient-based learning applied to document recognition. Proceedings of the IEEE 86, 2278-2324.

LeCun, Y., Cortes, C., and Burges, C.J.C. (1998b). The MNIST database of handwritten digits. (WWW), p. The MNIST database of handwritten digits.

Lerner, Y., Honey, C.J., Katkov, M., and Hasson, U. (2014). Temporal scaling of neural responses to compressed and dilated natural speech. J Neurophysiol 111, 2433-2444.

Li, N., Daie, K., Svoboda, K., and Druckmann, S. (2016). Robust neuronal dynamics in premotor cortex during motor planning. Nature advance online publication.

Lukoševičius, M., and Jaeger, H. (2009). Reservoir computing approaches to recurrent neural network training. Computer Science Review 3, 127-149.

Lyon, R.F. (1982). A computational model of filtering, detection, and compression in the cochlea. In Acoustics, Speech, and Signal Processing, IEEE International Conference on ICASSP '82, pp. 1282-1285.

Maass, W., Natschl, ger, T., and Markram, H. (2002). Real-Time Computing Without Stable States: A New Framework for Neural Computation Based on Perturbations. Neural Computation 14, 2531-2560.

Mante, V., Sussillo, D., Shenoy, K.V., and Newsome, W.T. (2013). Context-dependent computation by recurrent dynamics in prefrontal cortex. Nature 503, 78-84.

Mark Liberman, R.A., Ken Church, Ed Fox, Carole Hafner, Judy Klavans, Mitch Marcus, Bob Mercer, Jan Pedersen, Paul Roossin, Don Walker, Susan Warwick, Antonio Zampolli (1993). TI 46-Word LDC93S9. (Philadelphia, Linguistic Data Consortium).

Martens, J., and Sutskever, I. (2011). Learning recurrent neural networks with Hessian-free optimization. Proc 28th Int Conf Machine Learn.

Mauk, M.D., and Buonomano, D.V. (2004). The Neural Basis of Temporal Processing. Ann Rev Neurosci 27, 307-340.

Mello, G.B.M., Soares, S., and Paton, J.J. (2015). A scalable population code for time in the striatum. Curr Biol 9, 1113-1122.

Merchant, H., Harrington, D.L., and Meck, W.H. (2013). Neural Basis of the Perception and Estimation of Time. Annual Review of Neuroscience 36, 313-336.

Miller, J.L., Grosjean, F., and Lomanto, C. (1984). Articulation Rate and Its Variability in Spontaneous Speech - a Reanalysis and Some Implications. Phonetica 41, 215-225.

Mnih, V., Kavukcuoglu, K., Silver, D., Rusu, A.A., Veness, J., Bellemare, M.G., Graves, A., Riedmiller, M., Fidjeland, A.K., Ostrovski, G., *et al.* (2015). Human-level control through deep reinforcement learning. Nature 518, 529-533.

Murray, J.M., and Escola, S.G. (2017). Learning multiple variable-speed sequences in striatum via cortical tutoring. eLife 6, e26084.

Nobre, A.C., Correa, A., and Coull, J.T. (2007). The hazards of time. Current Opinion in Neurobiology 17, 465-470.

Ostojic, S. (2014). Two types of asynchronous activity in networks of excitatory and inhibitory spiking neurons. Nature neuroscience 17, 594-600.

Pearlmutter, B.A. (1995). Gradient calculation for dynamic recurrent neural networks: A survey. IEEE Trans on Neural Network 6, 1212–1228.

Rabiner, L. (1989). A tutorial on hidden Markov models and selected applications in speech recognition. Proceedings of the IEEE 77, 257-286.

Rabinovich, M., Huerta, R., and Laurent, G. (2008). Transient Dynamics for Neural Processing. Science 321, 48-50.

Rajan, K., Abbott, L., and Sompolinsky, H. (2010a). Inferring Stimulus Selectivity from the Spatial Structure of Neural Network Dynamics. Advances in Neural Information Processing Systems 23.

Rajan, K., and Abbott, L.F. (2006). Eigenvalue spectra of random matrices for neural networks. Physical review letters 97, 188104.

Rajan, K., Abbott, L.F., and Sompolinsky, H. (2010b). Stimulus-dependent suppression of chaos in recurrent neural networks. Physical Rev E 82, 011903(011905).

Rajan, K., Harvey, Christopher D., and Tank, David W. (2016). Recurrent Network Models of Sequence Generation and Memory. Neuron.

Schneider, D.M., and Mooney, R. (2015). Motor-related signals in the auditory system for listening and learning. Current Opinion in Neurobiology 33, 78-84.

Sebastián-Gallés, N., Dupoux, E., Costa, A., and Mehler, J. (2000). Adaptation to time-compressed speech: Phonological determinants. Perception & Psychophysics 62, 834-842.

Sompolinsky, H., Crisanti, A., and Sommers, H.J. (1988a). Chaos in random neural networks. Physical Rev Let 61, 259-262.

Sompolinsky, H., Crisanti, A., and Sommers, H.J. (1988b). Chaos in Random Neural Networks. Physical Review Letters 61, 259-262.

Sussillo, D., and Abbott, L.F. (2009). Generating Coherent Patterns of Activity from Chaotic Neural Networks. Neuron 63, 544-557.

Vogels, T.P., Sprekeler, H., Zenke, F., Clopath, C., and Gerstner, W. (2011). Inhibitory Plasticity Balances Excitation and Inhibition in Sensory Pathways and Memory Networks. Science 334, 1569-1573.

Waibel, A., Hanazawa, T., Hinton, G., Shikano, K., and Lang, K.J. (1989). Phoneme recognition using time-delay neural networks. Acoustics, Speech and Signal Processing, IEEE Transactions on 37, 328-339.

Wallace, E., Maei, H.R., and Latham, P.E. (2013). Randomly Connected Networks Have Short Temporal Memory. Neural computation 25, 1408-1439.

Wang, X.J. (2001). Synaptic reverberation underlying mnemonic persistent activity. Trends Neurosci 24, 455–463.

## Supplementary Information

**Supplementary Figures**

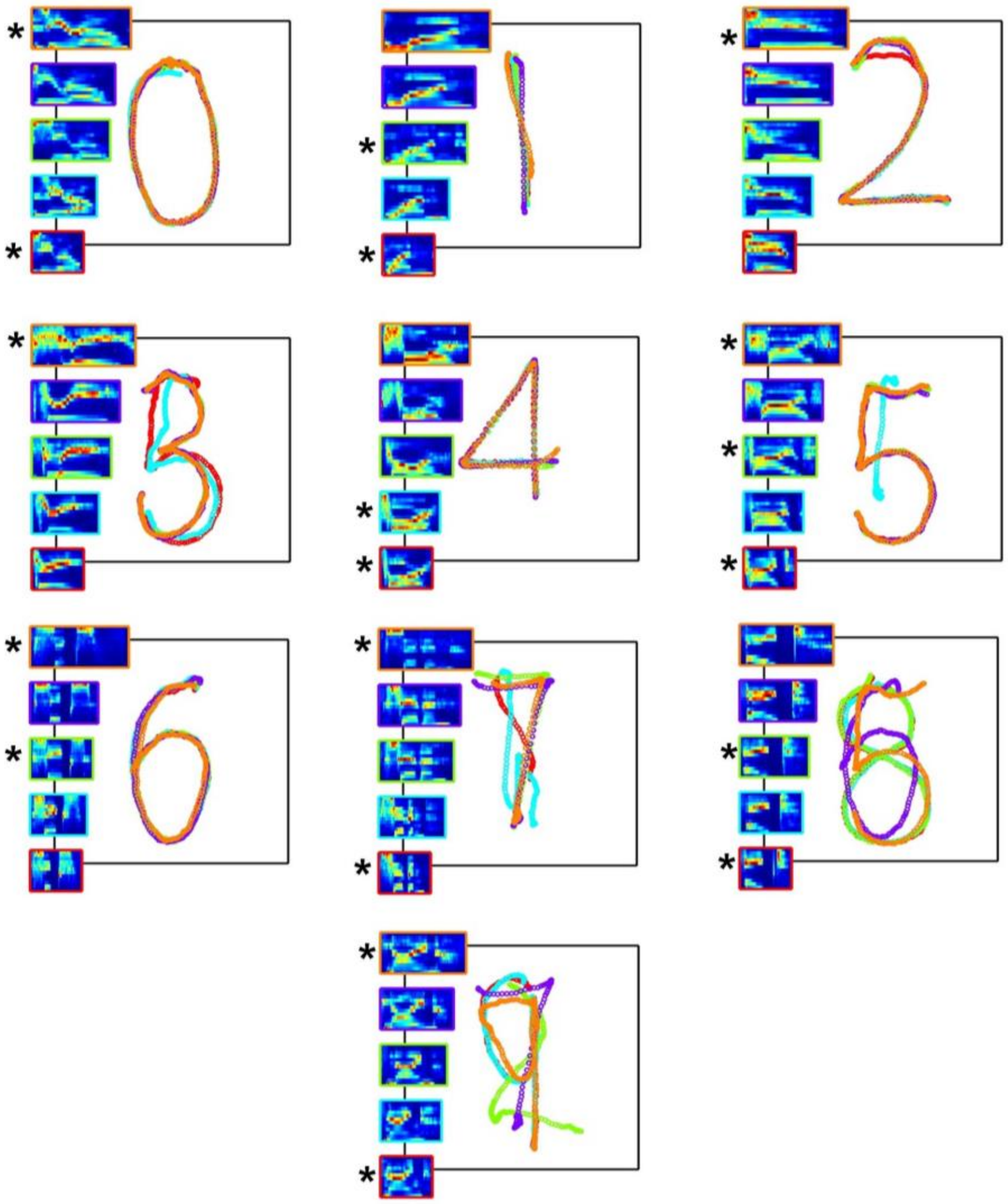

**Supplementary Figure 1: Transcription performance of an RNN trained only during the motor epoch.** Overlaid outputs of a motor-trained RNN (*N*=4000) for 5 sample utterances of each of the 10 digits used in the spoken-to-handwritten digit transcription task. Transcription performance of the motor-trained RNN is poorer than its sensorimotor trained counterpart (**Fig. 1B**). Each output pattern is color-coded to the bounding box of the corresponding cochleogram (insets). Sample utterances shown are a mix of trained (*) and novel utterances, and are the same ones shown in **Fig. 1B**. The network was trained on 90 utterances (10 digits, 3 subjects, 3 utterances per subject per digit). $I_0$ was set to 0.5 during network training, and to 0.05 during output training and testing.

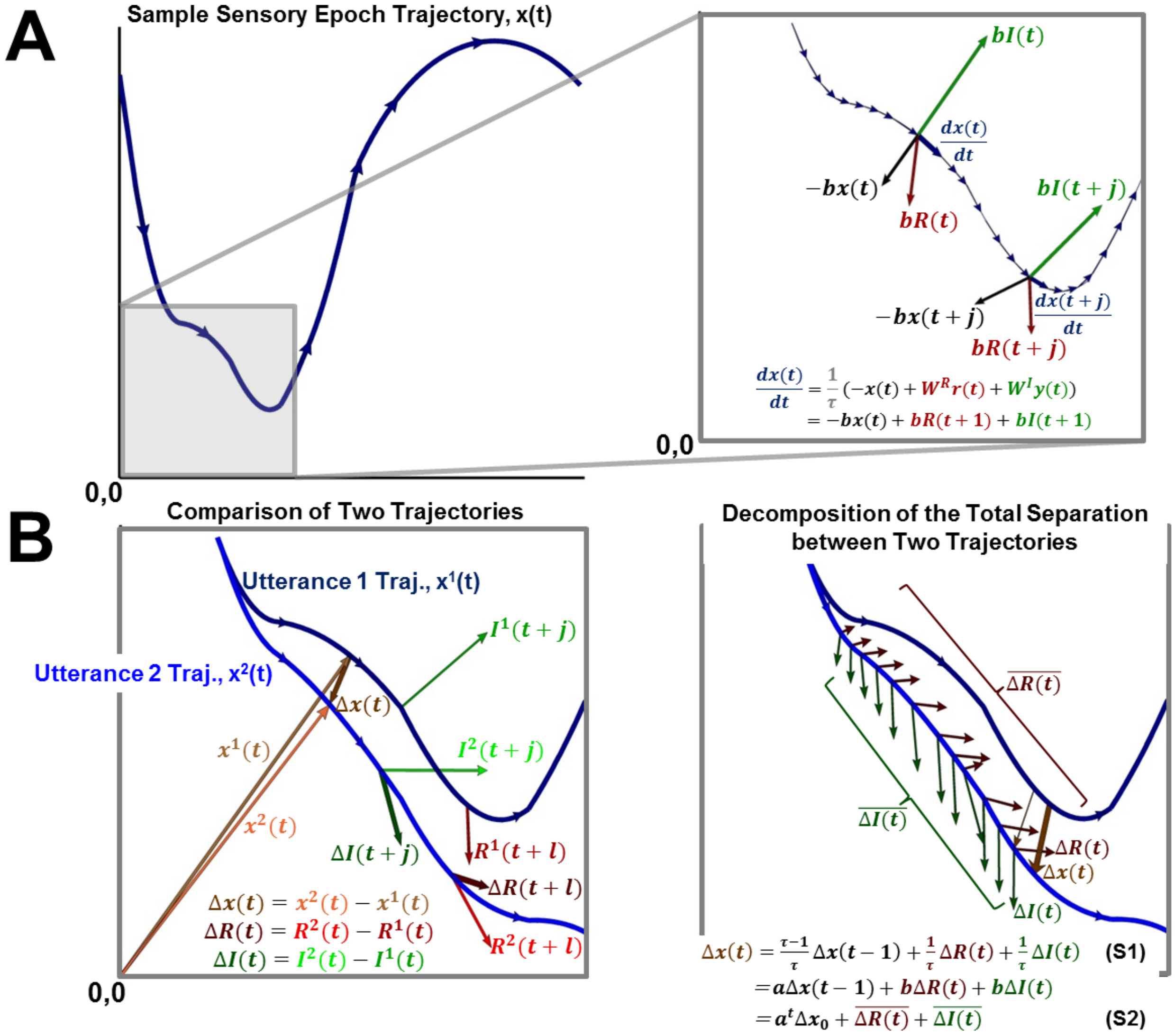

**Supplementary Figure 2: Schematic description of the decomposition of trajectories and trajectory separation into recurrent and input components.**

**(A)** Left panel. Schematic of the evolution of a trajectory *x(t)*. Right panel. The evolution of the trajectory in phase space between time steps *t* and *t+1* $\left(\frac{dx(t)}{dt}\right)$, can be decomposed into three component vectors: recurrent (red), input (green), and a passive component (black, which always points towards the origin). The sum of these vectors must equal the net movement of the trajectory between those time steps. **(B)** Left panel. The total deviation between two trajectories $\boldsymbol{x^1(t)}$ and $\boldsymbol{x^2(t)}$ at time step *t* (*Δx(t)*), can be viewed similarly as being composed of the recurrent (*ΔR(t)*), input (*ΔI(t)*) and total (*Δx(t-1)*) deviations at time *t-1*. Right panel. The total deviation at *t*, in turn, can be decomposed into the recurrent and input deviations at the previous time steps (i.e. Equation S1 may be viewed as a recurrence relationship). Thus, the total deviation at *t+1* can be decomposed into recurrent and input deviation histories ($\overline{\Delta R(t)}$ and $\overline{\Delta I(t)}$, respectively, where the bar represents an the exponentially weighted sum of the history up to time *t*). The vectors associated with ***ΔR*** and ***ΔI*** are plotted at different time points for visualization purposes.

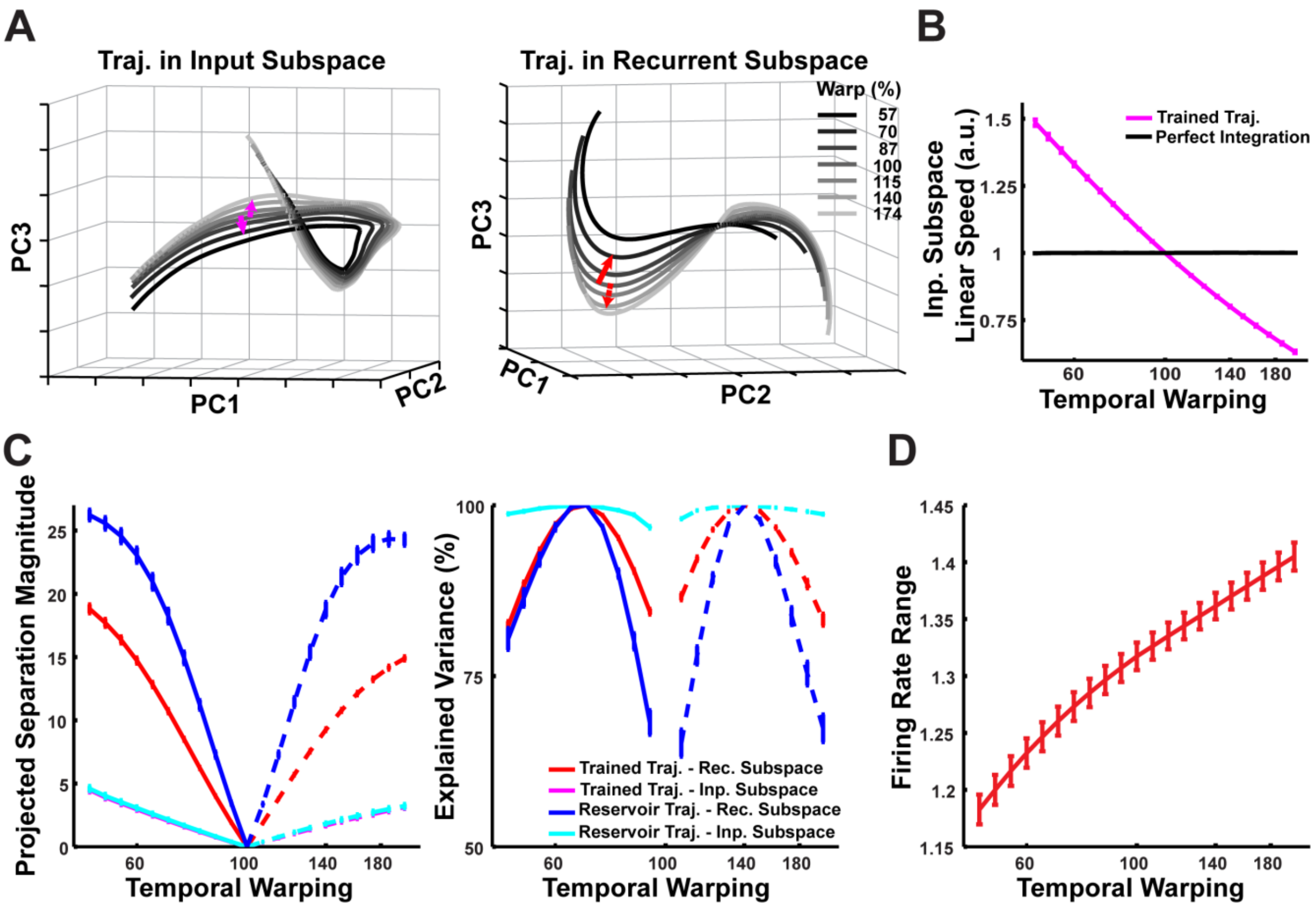

**Supplementary Figure 3: .Phase space relationships in the input and recurrent subspaces as a function of the warp factor.**

**(A)** Projections in PCA space of the input (left) and recurrent (right) subspace trajectories that compose the sensory epoch trajectories of a trained network (N=2100), in response to warped utterances of digit one, over a 100ms duration of the 1x warp trajectory and the corresponding utterance interval of the remaining trajectories. While the external inputs are originally inseparable in phase space (**Fig. 8C**), following integration by the RNN, these trajectories appear to separate out into parallel trajectories. A similar separation is observed in the recurrent subspace. **(B)** Time-averaged linear speed of sensory epoch trajectories in the input subspace of phase space where the inputs are leaky integrated (as in (A)), in comparison to the linear speed if the inputs were perfectly integrated. Whereas perfect integration produces constant speed trajectories (**Fig. 8B**), leaky integration instead results in trajectories that are a balance of constant speed and constant distance. The speeds are normalized to the 1x warp. Error bars indicate standard errors of the mean over the digits. **(C)** Time-averaged within-digit separation of reservoir and trained sensory epoch trajectories at each warp factor, from the respective 1x warp trajectory, along a common time-dependent direction (left), establishes that they are parallel. Following temporal alignment, at each time step, the separation between a test trajectory and the corresponding 1x warp trajectory is projected onto the direction of separation between the 0.7x and 1x warp trajectories (solid), or the 1.4x and 1x warp trajectories (dashed), depending on the input warp factor for the test trajectory (directions indicated by arrows in (A)). Monotonically increasing separation from the 1x warp as a function of the input warp factor establishes that the trajectories are parallel over the manifold. The separation along these manifolds explains a majority of the variance in the overall separation of trajectories from the 1x warp trajectory (right), although the explained variance is lowest for the recurrent dynamics of the reservoir owing to its partially chaotic dynamics. Error bars indicate standard errors of the mean over the digits. **(D)** Mean range of recurrent unit firing rates (defined as $\mathrm{mean}_i(\max_t(r_{i,t}) - \min_t(r_{i,t})))$ during sensory epoch trajectories of the trained network, over a range of warp factors. Error bars indicate standard errors of the mean over the digits. The network was trained with $I_0$ set to 0.25, and trajectories were recorded with $I_0$ set to 0.

## Supplementary Movies

### Supplementary Movie 1: A trained RNN performs the sensorimotor spoken-to-handwritten digit transcription task on novel utterances.

A trained RNN (N=4000) and its output units perform the transcription task on 5 novel utterances (5 different speakers). The last 2 utterances illustrate RNN performance on same-digit utterances of different durations (i.e. temporal invariance).Top Left: Input to the RNN (with audio). Moving gray bar indicates the current time step. Bottom Left: Evolution of RNN activity in a subset of the network (100 units). Right: Evolution of the output. The location of the pen is imprinted in red when the z co-ordinate is greater than 0.5, and plotted in gray otherwise.